\documentclass[aps,prd,floatfix,nofootinbib,superscriptaddress,preprint]{revtex4-1}
\usepackage{amssymb}
\usepackage{amsmath}
\usepackage{amsfonts}
\usepackage{graphicx}
\usepackage{color}
\usepackage{xspace}
\usepackage{ulem}
\usepackage{amsmath,amssymb,amsfonts,latexsym,cancel}
\usepackage{mathrsfs}
\usepackage{xcolor}                 
\usepackage{cancel}
\usepackage{anysize}
\usepackage{wrapfig}
\usepackage{slashed}
\usepackage{lipsum}
\usepackage{multirow}

\def \znbb {0\nu\beta\beta}

\def\ket#1{\left\vert #1\right\rangle}
\def\bra#1{\left\langle #1\right\vert}

\def\bf#1{\mathbf{#1}}

\newcommand{\AddrAHEP}{
AHEP Group, Instituto de F\'{\i}sica Corpuscular --
CSIC/Universitat de Val{\`e}ncia \\
Edificio de Institutos de Paterna, Apartado 22085,
E--46071 Val{\`e}ncia, Spain
}

\newcommand{\AddrUCL}{
Department of Physics and Astronomy, University College London,\\
London WC1E 6BT, United Kingdom
}

\newcommand{\AddrClermont}{
Laboratoire de Physique de Clermont, CNRS/IN2P3 – UMR 6533, Campus des C\'ezeaux,\\ 
4 Avenue Blaise Pascal, F-63178 Aubi\`ere Cedex, France
}

\newcommand{\AddrHEPHY}{
Institut f\"ur Hochenergiephysik, \"Osterreichische Akademie der Wissenschaften,\\
Nikolsdorfer Gasse 18, 1050 Wien, Austria
}

\newcommand{\AddrUTFSM}{
Department of Physics, Universidad T\'ecnica Federico Santa Mar\'ia,\\
Avenida Espa\~na 1680, Valpara\'iso, Chile
}

\def\suppmat{Appendix}

\begin{document}

\preprint{IFIC/18-38}

\title{Neutrinoless Double-$\beta$ Decay\\with Nonstandard Majoron Emission}

\author{Ricardo Cepedello} 
\email{ricepe@ific.uv.es}\affiliation{\AddrAHEP}\affiliation{\AddrUCL}
\author{Frank F. Deppisch} 
\email{f.deppisch@ucl.ac.uk}\affiliation{\AddrUCL}\affiliation{\AddrHEPHY}
\author{Lorena Gonz\'alez}
\email{lorena.gonzalez@alumnos.usm.cl}\affiliation{\AddrUTFSM}\affiliation{\AddrUCL}
\author{Chandan Hati} 
\email{chandan.hati@clermont.in2p3.fr}\affiliation{\AddrClermont}
\author{Martin Hirsch} 
\email{mahirsch@ific.uv.es}\affiliation{\AddrAHEP}

\begin{abstract}
\noindent 
We present a novel mode of neutrinoless double-$\beta$ decay with
emission of a light Majoron-like scalar particle $\phi$. We assume it
couples via an effective 7-dimensional operator with a $(V+A)$ lepton
current and $(V\pm A)$ quark currents leading to a long-range
contribution that is unsuppressed by the light neutrino mass. We
calculate the total double-$\beta$ decay rate and determine the fully
differential shape for this mode. We find that future double-$\beta$
decay searches are sensitive to scales of the order $\Lambda_\text{NP}
\approx 1$~TeV for the effective operator and a light scalar $m_\phi <
0.2$~MeV, based on ordinary double-$\beta$ decay Majoron searches. The
angular and energy distributions can deviate considerably from that of
two-neutrino double-$\beta$ decay, which is the main background. We point out
possible ultraviolet completions where such an effective
operator can emerge.
\end{abstract}

\maketitle

\section{Introduction}
\label{sec:intro}
double-$\beta$ decay processes are sensitive probes of physics beyond the Standard Model (SM). The SM process of two-neutrino double-$\beta$ ($2\nu\beta\beta$) decay is the rarest process ever observed with half lives of order $T_{1/2}^{2\nu\beta\beta} \sim 10^{21}~\text{y}$. Neutrinoless double-$\beta$ ($0\nu\beta\beta$) decay, with no observation of any missing energy, is clearly the most important mode beyond the SM as it probes the Majorana nature and mass $m_\nu$ of light neutrinos, with current experiments sensitive as $T_{1/2}^{0\nu\beta\beta} \sim (0.1~\text{eV}/m_\nu)^2 \times 10^{26}~\text{y}$. In general, it is a crucial test for any new physics scenario that violates lepton number by two units.

On the other hand, one or more exotic neutral particles may also be emitted, with a signature of anomalous missing energy beyond that expected in $2\nu\beta\beta$ decay. A well studied set of theories involve the emission of a scalar particle, called \textit{Majoron} $J$. The first such proposed Majoron was a Goldstone boson associated with the spontaneous breaking of lepton number symmetry \cite{Chikashige:1980ui, Gelmini:1980re}, coupling to a neutrino $\nu$ as $g_J\nu\nu J$, cf. Fig.~\ref{fig:diagram}~(left). Current searches have a sensitivity of the order $T_{1/2}^{0\nu\beta\beta J} \sim (10^{-5}/g_J)^2 \times 10^{24}~\text{y}$. The term Majoron has been used in a wider sense, implying just a charge-neutral scalar particle (Goldstone boson or not) or vector particle \cite{Carone:1993jv}. Originally considered to be massless, it may also be a light particle \cite{Bamert:1994hb, Hirsch:1995in, Blum:2018ljv} that can potentially be a Dark Matter candidate \cite{Berezinsky:1993fm, Garcia-Cely:2017oco, Brune:2018sab}. Searches for extra particles in double-$\beta$ decay are crucial in understanding neutrinos. Most importantly, violation of lepton number by two units and thus the Majorana nature of neutrinos can only be firmly established in the case of $0\nu\beta\beta$ decay.

Not all such emission modes have been discussed in the literature. Existing experimental searches so far focus on the emission of one or two Majorons originating from the intermediate neutrino exchanged in the process. The different Majoron scenarios have been classified into several categories, all of which assume SM $(V-A)$ charged currents with the electrons and quarks. In this Letter, we instead consider $0\nu\beta\beta\phi$ decay with emission of a light neutral scalar $\phi$ from a single effective dimension-7 operator of the form $\Lambda_\text{NP}^{-3}(\bar u\mathcal{O} d)(\bar e\mathcal{O}\nu)\phi$, cf. Fig.~\ref{fig:diagram}~(center), with the fermion currents having a different chiral structure from that in the SM. In the following, we will refer to the light scalar as ``Majoron'', independent of its origin. We determine the sensitivity to $\Lambda_\text{NP}$ and analyse the effect on the energy and angular distributions in comparison with $2\nu\beta\beta$ decay. We also comment on ultraviolet scenarios underlying the effective operator.

\section{Effective Long-Range Interactions}
\label{sec:eft}
\begin{figure}[t!]
	\centering
	\includegraphics[scale=0.9]{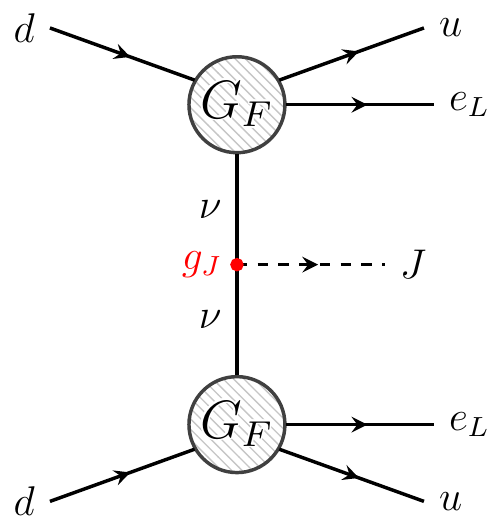}
	\includegraphics[scale=0.9]{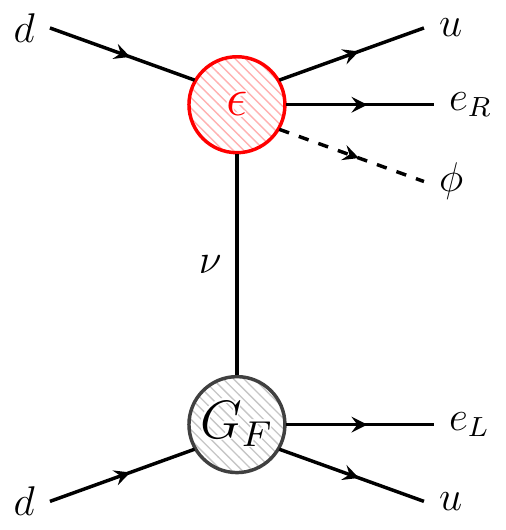}
	\includegraphics[scale=0.9]{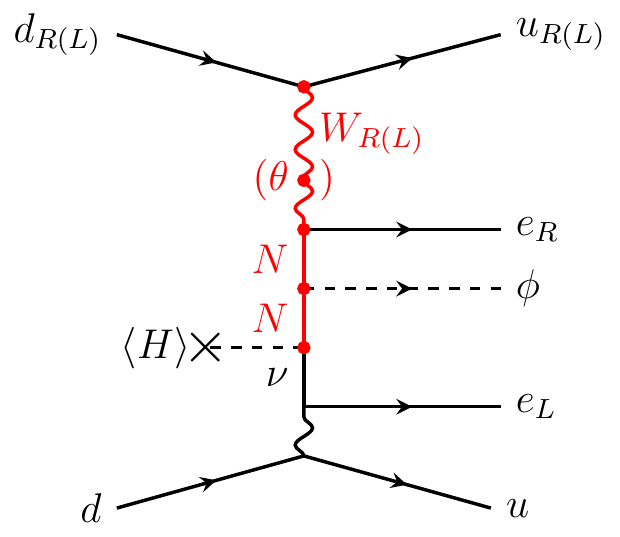}
	\caption{Feynman diagrams for ordinary $0\nu\beta\beta J$ Majoron decay (left), $0\nu\beta\beta\phi$ decay triggered by an effective operator of the form $\Lambda^{-3}_\text{NP}(\bar u\mathcal{O} d)(\bar e\mathcal{O}\nu)\phi$ (center) and possible ultraviolet completion of the latter in a Left-Right symmetric model (right).}
	\label{fig:diagram} 
\end{figure}
We are interested in processes where right- and left-handed electrons are emitted along with a scalar $\phi$ considering as a first approach, only $(V+A)$ and $(V-A)$ currents. The effective Lagrangian can then be written as
\begin{align}
\label{eq:lagrangian}
	\mathcal{L}_{0\nu\beta\beta\phi} &=
	\frac{G_F \cos\theta_C}{\sqrt{2}}\left(
		  j^\mu_L J^{\phantom{\mu}}_{L\mu} 
		+ \frac{\epsilon^\phi_{RL}}{m_p} j_R^\mu J^{\phantom{\mu}}_{L\mu}\phi 
		+ \frac{\epsilon^\phi_{RR}}{m_p} j_R^\mu J^{\phantom{\mu}}_{R\mu}\phi
	\right) + \text{h.c.},
\end{align}
with the Fermi constant $G_F$, the Cabbibo angle $\theta_C$, and the leptonic and hadronic currents $j_{L,R}^\mu = \bar e\gamma^\mu(1\mp\gamma_5)\nu$ and $J^\mu_{L,R} = \bar u\gamma^\mu(1\mp\gamma_5)d$, respectively. Here, $\nu$ is a 4-spinor field of the light electron neutrino, either defined by $\nu = \nu_L + \nu_L^c$ (i.e. a Majorana spinor constructed from the SM active left-handed neutrino $\nu_L$) or $\nu = \nu_L + \nu_R$ (a Dirac spinor constructed from the SM $\nu_L$ and a new SM-sterile right-handed neutrino $\nu_R$). Whether the light neutrinos are of Majorana or Dirac type and whether total lepton number is broken or conserved is of crucial importance for an underlying model (determined by the chosen lepton numbers for $\nu_R$ and $\phi$) but as far as the effective interactions in Eq.~\eqref{eq:lagrangian} are concerned, this does not play a role in our calculations. The proton mass $m_p$ is introduced in the exotic interactions as normalization to make the effective coupling constants $\epsilon^\phi_{RL}$ and $\epsilon^\phi_{RR}$ dimensionless, in analogy to the effective operator treatment of $0\nu\beta\beta$ decay \cite{Pas:1999fc,Deppisch:2012nb}. In Eq.~\eqref{eq:lagrangian}, we omit exotic operators with left-handed lepton currents; as in the standard long-range case, such contributions will be additionally suppressed by the small neutrino masses \cite{Deppisch:2012nb}. We instead focus on the process depicted in Fig.~\ref{fig:diagram}~(center), where the SM $(V-A)$ Fermi interaction, the first term in Eq.~\eqref{eq:lagrangian}, meets one of the exotic operators. In this case, the momentum part in the numerator of the neutrino propagator contributes, rather than the mass. In Eq.~\eqref{eq:lagrangian} we consider the first generation electron and neutrino only. Generalizing to three flavors amounts to promoting the $\epsilon^\phi_{RX}$ couplings to $3\times 3$ matrices in generation space, $(\epsilon^\phi_{RX})_{\alpha i}$ ($\alpha = e, \mu, \tau$, $i = \nu_1, \nu_2, \nu_3$). The final decay rate will then be proportional to $|\epsilon^\phi_{RX}|^2 \to |\sum_i (\epsilon^\phi_{RX})_{ei} U_{ei}|^2$, where $U$ is the SM lepton mixing matrix.

\section{Decay Rate and Distributions}
\label{sec:calculation}

We base our calculation of the $0\nu\beta\beta\phi$ decay rate and
kinematic distributions on Doi et al.~\cite{Doi:1985dx}. The
details of the calculation are given in the \suppmat{}~A; here we 
outline the main features. Summing over all intermediate nuclear
states $N$, the amplitude of $0_I^+ \to 0_F^+$~$0\nu\beta\beta\phi$
decay can be written as
\begin{align}
\label{eq:amplitude}
    \mathcal{M} &= \epsilon^\phi_{RX}\frac{(G_F\cos\theta_C)^2}{\sqrt{2}m_p}
    \sum_N \int d^3x d^3y \int\frac{d^3q}{2\pi^2\omega} \phi(\bf{y}) e^{i \bf{q}(\bf{x}-\bf{y})}  \nonumber\\ 
    &\times\left\lbrace\!
        \Bigg[ 
        	  \frac{J_{LX}^{\rho\sigma}(\bf{x},\bf{y}) 
        	  	    u_{\rho\sigma}^L(E_1\bf{x},E_2\bf{y})}
        	       {\omega + \mu_N - \frac{1}{2} (E_1 - E_2 - E_\phi)} 
        \!-\! \frac{J_{XL}^{\rho\sigma}(\bf{x},\bf{y})
        	        u_{\rho\sigma}^R(E_1\bf{x},E_2\bf{y})}
        	       {\omega + \mu_N - \frac{1}{2} (E_1 - E_2 + E_\phi)} 
        \Bigg] 
        \!-\!  \Bigg[ E_1 \leftrightarrow E_2\Bigg]
    \!\right\rbrace.
\end{align}
Here, $X = L$, $R$ correspond to $\epsilon^\phi_{RL}$,
$\epsilon^\phi_{RR}$, $\mu_N = E_N - E_I + Q_{\beta\beta}/2 + m_e$
with $E_I$ and $E_N$ the energies of the initial and intermediate
nucleus, respectively. The energies of the two outgoing electrons and
the Majoron are $E_{1,2}$ and $E_\phi$, respectively, and the available kinetic energy release $Q_{\beta\beta}$. The nucleon and
lepton currents are defined as
\begin{align}
\label{eq:def J}
    J_{YX}^{\rho\sigma}(\bf{x},\bf{y}) 
    &= \bra{F} J_Y^\rho(\bf{x}) \ket{N} \bra{N} J_X^\sigma(\bf{y}) \ket{I},
    \quad X,Y = L,R,\\
    u_{\rho\sigma}^{L,R}(E_1\bf{x},E_2\bf{y}) 
    &= \frac{1}{2} q^\mu \bar e (E_1\bf{x}) \gamma_\rho \gamma_\mu \gamma_\sigma (1\mp\gamma_5) \, e^c(E_2\bf{y}).
\label{eq:def u}
\end{align}
We consider that the internal neutrino propagates between the
interaction points $\bf{x}$ and $\bf{y}$ with momentum
$q^\mu = (\omega,\bf{q})$. From Eqs.~\eqref{eq:amplitude} and
\eqref{eq:def u} one can see explicitly the required antisymmetry of
the amplitude under the exchange of the two electrons.

In Eq.~\eqref{eq:amplitude}, the Majoron energy $E_\phi$ is added or
subtracted depending on whether the electron labeled 1 or 2 is
being emitted from the exotic operator. The
Majoron makes a crucial difference, as $E_\phi$ goes together with $(E_1 -
E_2)$ and not with the term proportional to the intermediate nuclei
energy $\mu_N$ as for an ordinary Majoron. A dependence on $E_\phi$ will thus appear through the matrix element in
addition to that through the phase space.
The differential decay rate for the $0^+\to 0^+$ $0\nu\beta\beta\phi$
decay can then be written as \cite{Doi:1985dx}
\begin{align}
\label{eq:decay rate}
	d\Gamma = \frac{(G_F\cos\theta_C g_A)^4 m_e^2}{256\pi^7 (m_p R)^2}
	\left[a(E_1,E_2) + b(E_1,E_2) \cos\theta\right] 
	p_1 p_2 E_1 E_2 E_\phi dE_1 dE_2 d\!\cos\theta,
\end{align}
with the axial coupling $g_A$ of the nucleon and the radius $R$ of the
nucleus. The magnitudes of the electron spatial momenta are denoted
$p_{1,2}$ and $0 \leq \theta \leq \pi$ is the angle between the
emitted electrons. In Eq.~\eqref{eq:decay rate}, the Majoron energy is
determined as $E_\phi = Q_{\beta\beta} + 2m_e - E_1 - E_2$ by energy
conservation. Definitions for the coefficients $a(E_1, E_2)$, $b(E_1, E_2)$ in the decay rate can be found in the \suppmat{}~A, where we show in detail the derivation of the differential decay rate. Therein we use the nuclear matrix elements listed in Table~I and Coulomb-corrected relativistic electron wave functions.

\begin{figure}
	\centering
	\includegraphics[clip,trim={10 10 0 0},width=0.49\textwidth]{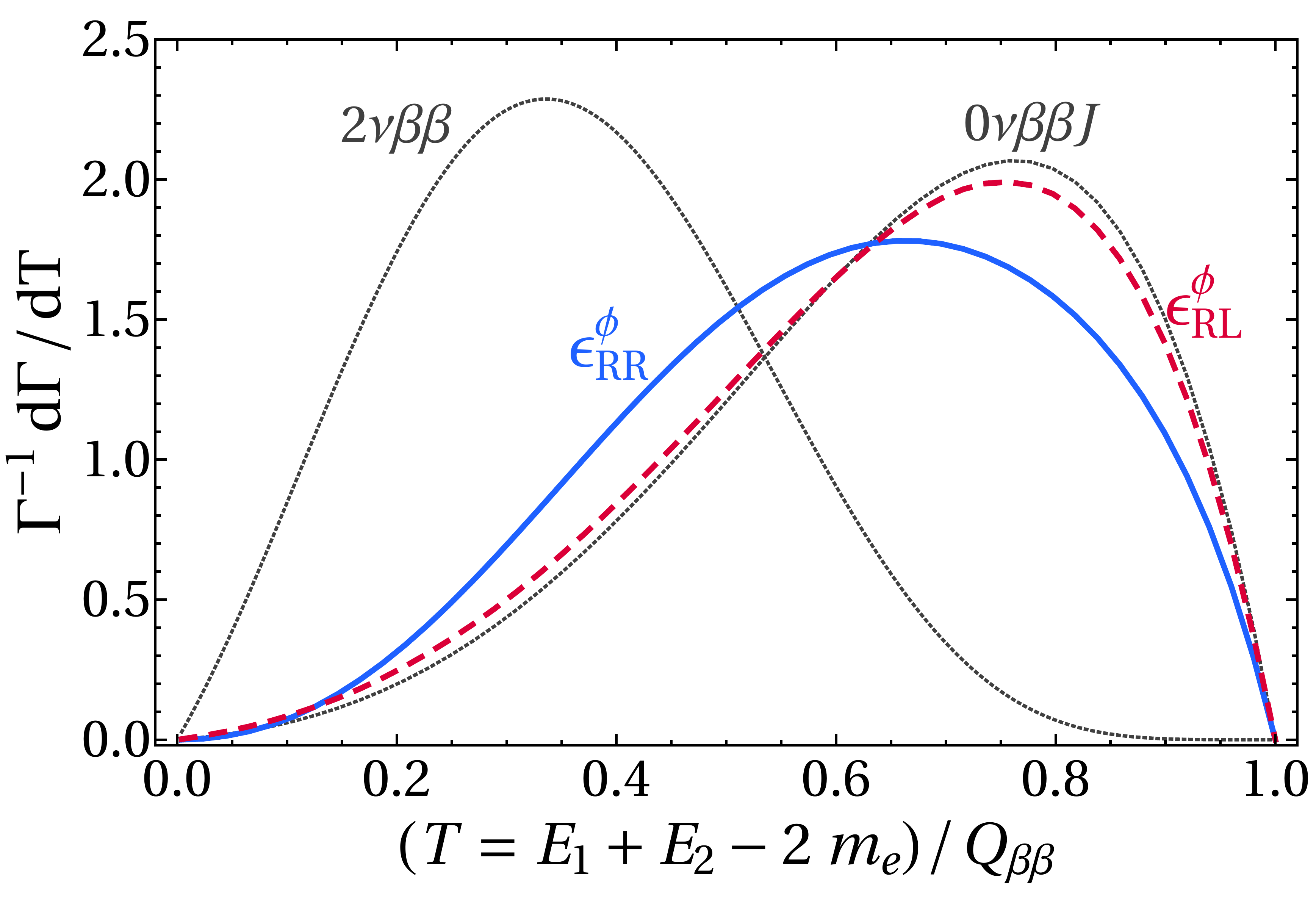}
	\includegraphics[clip,trim={10 10 0 0},width=0.49\textwidth]{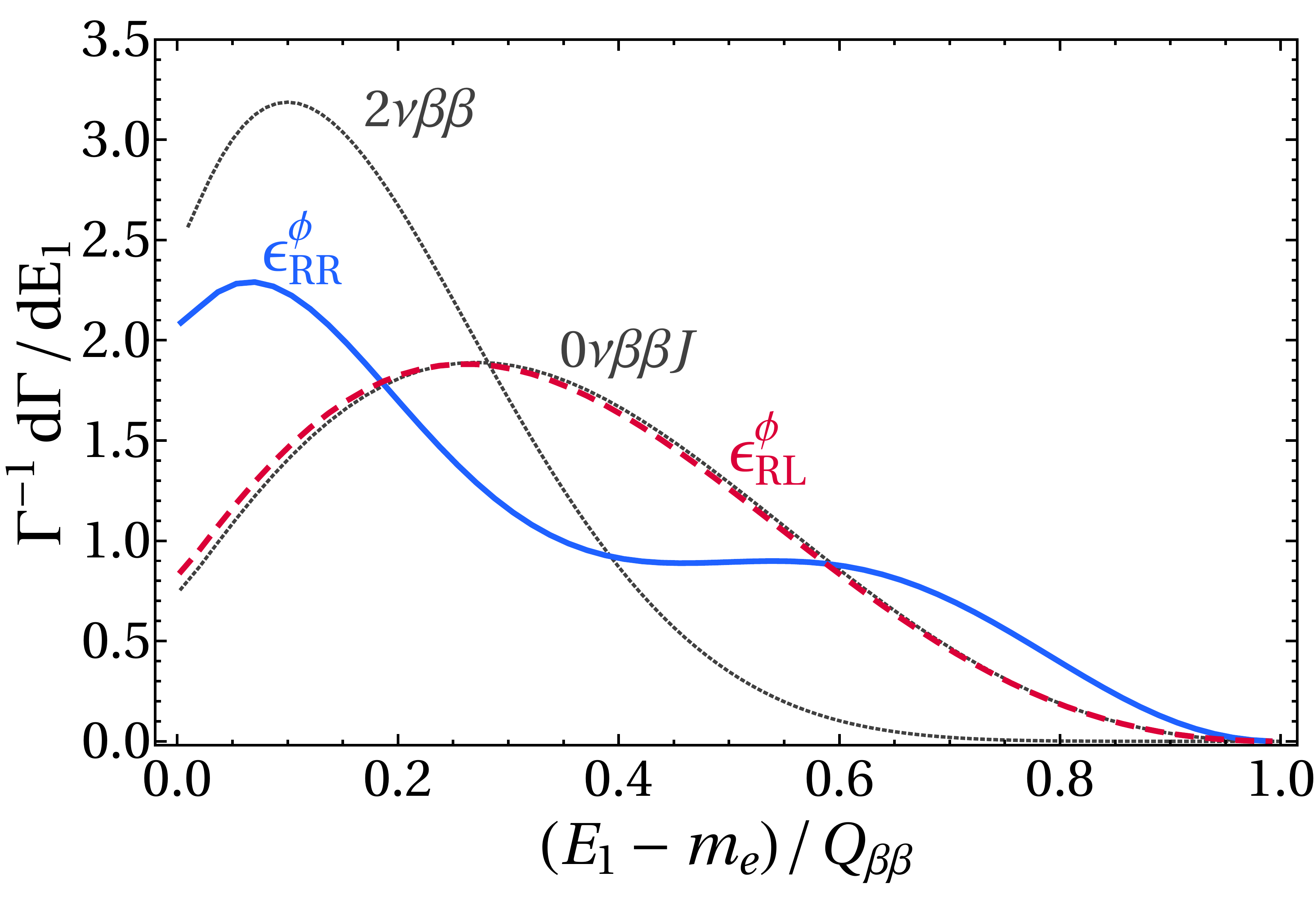}
	\caption{Left: Normalized $0\nu\beta\beta\phi$ decay distributions in the total kinetic energy of the electrons for $^{136}\text{Xe}$. Right: Normalized $0\nu\beta\beta\phi$ decay distribution in the single electron kinetic energy distribution for $^{82}\text{Se}$. The blue solid and red dashed lines correspond to the $\epsilon^\phi_{RR}$ and $\epsilon^\phi_{RL}$ cases, respectively. The corresponding distributions for the SM $2\nu\beta\beta$ decay and ordinary $0\nu\beta\beta J$ Majoron decay (spectral index $n = 1$) are given for comparison.}
	\label{fig:energydist}
\end{figure}
From Eq.~\eqref{eq:decay rate}, the total decay rate and thus the half life is calculated by performing the integration of $a(E_1,E_2)$ over all energies within the allowed phase space limits $E_1$, $E_2 \geq 0$ and $E_1 + E_2 \leq Q_{\beta\beta} + 2m_e$. In addition, we determine and discuss several distributions below. We will show results for the $\epsilon^\phi_{RL}$ and $\epsilon^\phi_{RR}$ versions of the effective operators where we consider only one of these to be present at a time. We assume the exotic $\phi$ Majoron to be massless in our calculations and comment on massive $\phi$ in the discussion below. For our numerical evaluation we focus on two isotopes: (i) $^{136}\text{Xe}$, for which the KamLAND-Zen collaboration \cite{KamLAND-Zen:2016pfg} currently provides the most stringent constraints; (ii) $^{82}\text{Se}$ used by NEMO-3 and the upcoming SuperNEMO experiments \cite{Arnold:2018tmo} that can measure the detailed electron topology.

For all experimental searches, the crucial distribution is with respect to the sum of the kinetic energies of the detected electrons. With the SM $2\nu\beta\beta$ decay as irreducible background to any exotic signal, it is important to calculate it precisely. In Fig.~\ref{fig:energydist}~(left), we compare the normalized total electron kinetic energy distribution of $0\nu\beta\beta\phi$ decay with that of $2\nu\beta\beta$ decay and ordinary $0\nu\beta\beta J$ Majoron decay (with spectral index $n = 1$) for the isotope $^{136}\text{Xe}$. The distribution associated with $\epsilon^\phi_{RL}$ is very similar to ordinary $0\nu\beta\beta J$ decay, while the introduction of a hadronic right-handed current in the $\epsilon^\phi_{RR}$ term changes considerably the shape of the distribution. In both cases, the spectral index still corresponds to $n=1$ with the characteristic onset near the kinematic endpoint. We emphasize that because of the different shape, a dedicated signal over background analysis is required to determine the experimental sensitivity on the effective parameters $\epsilon^\phi_{RL}$ and $\epsilon^\phi_{RR}$ precisely.

NEMO-3 and SuperNEMO are able to measure the individual electron energies. In right-handed current scenarios without emission of a Majoron, the single energy distribution exhibits a distinctive valley-type shape. This occurs as the dominant term is proportional to $(E_1 - E_2)$ for the corresponding $\epsilon_{RR}$ term, as a result of the antisymmetry with respect to electron exchange.\footnote{For the $\epsilon_{RL}$ term with a left-handed hadronic current, $P$-wave and nuclear recoil contribute constructively, giving a dominant contribution proportional to $(E_1 + E_2)$ \cite{Doi:1985dx}.} In our case, depicted in Fig.~\ref{fig:energydist}~(right), part of the energy is being carried away by the Majoron, shifting the distribution towards lower electron energies and softening the characteristic valley-type distribution for $\epsilon^\phi_{RR}$. The distribution does not vanish for $E_1 - m_e = \frac{1}{2}Q_{\beta\beta}$ (as in the ordinary right-handed current case), but is still significantly different from that of ordinary Majoron emission. The distribution with respect to both electron energies is depicted in Fig.~\ref{fig:angular2}~(top panel) in the \suppmat{}~A. It exhibits an even more pronounced difference between the $\epsilon^\phi_{RR}$ mode and $2\nu\beta\beta$. This may be used experimentally to improve the sensitivity through kinematic selection criteria, counteracting the effect of the less peaked total energy distribution, cf. Fig.~\ref{fig:energydist}~(left).

One can use also angular correlations to distinguish between
left-handed and right-handed currents \cite{Doi:1983wv,Arnold:2010tu},
see Fig.~\ref{fig:angular2}~(bottom panel) in the \suppmat{}~A.
Integrating over the electron energies one obtains the average angular
distribution which takes the simple form
$\frac{d\Gamma}{d\!\cos\theta} = \frac{\Gamma}{2}(1 +k\cos\theta)$.
The coefficient $k$ is $k^\phi_{RL} = + 0.70$ (electrons are dominantly emitted collinearly) and $k^\phi_{RR} = -0.05$ (electrons are emitted nearly isotropically)
in our $0\nu\beta\beta\phi$ scenarios with $\epsilon^\phi_{RL}$ and
$\epsilon^\phi_{RR}$, respectively, for $^{82}\text{Se}$. For comparison, the angular correlation factor for SM
$2\nu\beta\beta$ decay is $k^{2\nu\beta\beta} = -0.66$ and $k^J =
-0.80$ for ordinary Majoron emission; i.e., the electrons are dominantly emitted back-to-back.

\setlength{\tabcolsep}{6pt}
\begin{table}
	\centering
	\begin{tabular}{cccc}
		\hline
		Isotope & $T_{1/2}$~[y] & $|\epsilon^\phi_{RL}|$ & $|\epsilon^\phi_{RR}|$ \\
        \hline
		{$^{82}\text{Se}$} & $3.7\times 10^{22}$ \cite{Arnold:2018tmo} & $ 4.1\times 10^{-4}$ & $4.6\times 10^{-2}$ \\
	    {$^{136}\text{Xe}$} & $2.6 \times 10^{24}$ \cite{KamLAND-Zen:2016pfg} & $1.1\times 10^{-4}$ & $1.1\times 10^{-2}$ \\
		\hline
		{$^{82}\text{Se}$} & $1.0\times 10^{24}$ & $8.0\times 10^{-5}$ & $8.8\times 10^{-3}$ \\
	    {$^{136}\text{Xe}$} & $1.0\times 10^{25}$ & $5.7\times 10^{-5}$ & $5.8\times 10^{-3}$ \\
		\hline
	\end{tabular}
\caption{Current limits and expected future sensitivity on the effective couplings $\epsilon^\phi_{RL}$ and $\epsilon^\phi_{RR}$ of $0\nu\beta\beta\phi$ decay for $^{82}\text{Se}$ and $^{136}\text{Xe}$. The limits are estimated based on the experimental half life constraints for ordinary Majoron emission (spectral index $n = 1$) as given. Nuclear matrix elements from Refs.~\cite{Horoi:2015gdv, Caurier:1996bu, Caurier:2007qn} where used for this estimate.}
\label{table:limits}
\end{table}
Finally, we estimate the sensitivity of existing and planned future double-$\beta$ decay searches on the effective coupling strength $\epsilon^\phi_{RL}$ and $\epsilon^\phi_{RR}$ of $0\nu\beta\beta\phi$ decay. We would like to emphasize again that due to the different total electron energy distribution, a dedicated signal over background analysis is required to determine the constraints precisely. Experiments such as NEMO-3 and SuperNEMO can also improve their sensitivity due to the non-standard decay topology, especially for $\epsilon^\phi_{RR}$. As detailed in the \suppmat{}~A, a requirement that any one electron has a kinetic energy of $E_i - m_e > Q_{\beta\beta}/2$ can for example reduce the $2\nu\beta\beta$ background by an order of magnitude. Here, we simply estimate the sensitivity by comparing our predictions for the $0\nu\beta\beta\phi$ decay half life $T_{1/2} = \ln 2/\Gamma$ with the experimental constraints on ordinary ($n=1$) Majoron emission. We use the most stringent limits for $^{82}\text{Se}$ by NEMO-3 \cite{Arnold:2018tmo} and for $^{136}\text{Xe}$ by KamLAND-Zen \cite{KamLAND-Zen:2016pfg}. For future prospects, we estimate that experimental Majoron search sensitivities may reach $T_{1/2}^\text{Se} \approx 10^{24}$~y (e.g. with the help of angular and energy selection cuts at SuperNEMO) and $T_{1/2}^\text{Xe} \approx 10^{25}$~y.\footnote{The corresponding $0\nu\beta\beta$ decay sensitivities of the planned SuperNEMO \cite{Macolino:2017vyd} and nEXO experiments \cite{gratta_giorgio_2018_1286892} may improve by $\mathcal{O}(100)$, but this requires an experimental approach that is essentially background-free. This is not possible for Majoron emission with a continuous total electron energy spectrum.} The corresponding limits on $\epsilon^\phi_{RL}$ and $\epsilon^\phi_{RR}$ are shown in Table~\ref{table:limits}, where only one effective operator is assumed to present at a time.

\section{Discussion}
\label{sec:discussion}

Searches for Majorons or Majoron-like particles are a staple in double
beta decay experiments. So far, they only cover the case where the
neutrino involved couples via the SM $(V-A)$ charged current
interaction. This is clearly a well-motivated minimal choice but it is
worthwhile to explore other scenarios. In this Letter, we have
discussed one such alternative where a Majoron-like particle $\phi$ is
emitted from effective operators with $(V+A)$ leptonic currents,
cf. Fig~\ref{fig:diagram}~(center). The future sensitivities on the
effective couplings $\epsilon^\phi_{RL}$ and $\epsilon^\phi_{RR}$ shown in Table~\ref{table:limits} may
be translated into effective operator scales $\Lambda_\text{NP}
\approx 1.3$~TeV and 270~GeV, respectively, using
$1/\Lambda_\text{NP}^{3} = \epsilon^\phi_{RX}
G_F\cos\theta_C/(\sqrt{2}m_p)$. As noted before, we assume a massless
$\phi$ in deriving these limits; they remain essentially unchanged for
masses small compared to $Q_{\beta\beta}$, $m_\phi \lesssim
0.2$~MeV and are of the same order for $m_\phi \lesssim 1$~MeV, but
will deteriorate as $m_\phi \to Q_{\beta\beta}$ (for a recent analysis in ordinary Majoron emission, see Ref.~\cite{Brune:2018sab}). Constraints on our operators may also be set from other processes, such as exotic decay modes of the pion, $\pi^- \to e^- \bar\nu_e \phi$. As we consider only $V+A$ currents, helicity suppression will still apply and the limits are expected to be correspondingly weak, we roughly estimate $\Lambda_\text{NP} \gtrsim 15$~GeV.

An ultraviolet scenario generating the effective operators in Eq.~\eqref{eq:lagrangian} is suggested in Left-Right symmetric models \cite{Pati:1974yy, Mohapatra:1974hk, Mohapatra:1974gc, Senjanovic:1975rk} where the SM $W$ and $\nu$ are replaced by their right-handed counterparts $W_R$ and $N$. The heavy neutrino $N$ then mixes with $\nu$ via a Yukawa coupling $y_\nu$ once the SM Higgs boson acquires its vacuum expectation value $\langle H \rangle = 174$~GeV. A massless or light scalar $\phi$ is not part of the minimal Left-Right symmetric model which thus needs to be modified, e.g. by keeping the $U(1)_{B-L}$ symmetry global or by extending its scalar sector. Charging $\phi$ under lepton number allows coupling to $N$ with a strength $y_N$. The corresponding diagram is shown in Fig.~\ref{fig:diagram}~(right). We can then identify
\begin{align}
	\frac{G_F \cos\theta_C}{\sqrt{2} m_p} \epsilon^\phi_{RR}
	= \frac{g_R^2 y_N y_\nu \langle H \rangle\cos\theta_C^R}{8m_{W_R}^2 m_N^2},
\end{align} 
leading to the estimate
\begin{align}
	\frac{T_{1/2}^\text{Xe}}{10^{25}~\text{y}} \approx
	\left(\frac{3.5\times 10^{-4}}{g_R^2 y_N y_\nu\cos\theta_C^R}\right)^2
	\left(\frac{m_{W_R}}{4~\text{TeV}}\right)^4
	\left(\frac{m_N}{100~\text{MeV}}\right)^4,
\end{align} 
where $g_R$ is the gauge coupling constant and $\theta_C^R$ the equivalent of the Cabibbo angle, both associated with the $SU(2)_R$ of the Left-Right symmetric model. Alternatively, it is also
possible to trigger the $\epsilon^\phi_{RL}$ mode through the
$W_R$-$W$ mixing $\theta$. Its value is generically expected to be $\theta = \kappa g_R m_W^2/(g_L m_{W_R}^2)$ where $\kappa = \mathcal{O}(1)$. In this case one has
\begin{align}
	\frac{G_F \cos\theta_C}{\sqrt{2} m_p} \epsilon^\phi_{RL}
	= \frac{g_R g_L\theta y_N y_\nu \langle H \rangle\cos\theta_C}{8 m_W^2 m_N^2},
\end{align} 
resulting in the estimate
\begin{align}
	\frac{T_{1/2}^\text{Xe}}{10^{25}~\text{y}} \approx
	\left(\frac{1.4\times 10^{-4}}{g_R^2 \kappa y_N y_\nu}\right)^2
	\left(\frac{m_{W_R}}{25~\text{TeV}}\right)^4
	\left(\frac{m_N}{100~\text{MeV}}\right)^4.
\end{align} 
This is more stringent due to the better sensitivity on $\epsilon^\phi_{RL}$ in Table~\ref{table:limits}. Choosing the right-handed neutrino mass $m_N$ to be as low as 100~MeV is strictly speaking not allowed in the effective operator treatment which requires $m_N \gg p_F \approx 100$~MeV, but it may be more natural in a scenario where the mass of $N$ is generated through the vacuum expectation value of $\phi$, $m_N = y_N\langle\phi\rangle$. In fact, choosing $m_N$ to be smaller and abandoning the effective operator treatment may be more natural; the qualitative arguments should hold as above though a dedicated calculation of $0\nu\beta\beta\phi$ would be required. In addition, the contribution to $0\nu\beta\beta\phi$ via a heavy neutrino is expected to peak at $m_N \approx p_F$ with the above estimates applying to a good approximation \cite{Simkovic:1999re}.\footnote{In addition to the operators discussed here, the Left-Right symmetric scenario will also induce a standard Majoron interaction $\phi\nu\nu$ (leading to standard Majoron emission with spectral index $n = 1$) after electroweak symmetry breaking from an operator of the form $\phi H H \nu\nu$. It is suppressed relative to our contributions by an additional power of $y_\nu$ but does not suffer from suppression by the heavy $W_R$ mass or the small $W_R-W$ mixing.} We can here only give a sketch of what such a model scenario may look like; for a more detailed discussion, we refer the reader to our \suppmat{}~B.1 where we describe a Left-Right symmetric model incorporating a Dirac seesaw mechanism to generate the light neutrino masses. Here, ordinary $0\nu\beta\beta$ decay is not allowed and only our $0\nu\beta\beta\phi$ mode would occur.

Other ultraviolet completions do exist; to lowest dimension, the effective operator $\epsilon^\phi_{RR}$ in Eq.~\eqref{eq:lagrangian} can be matched to the SM invariant operator $L e_R \bar d_R u_R H \phi$ ($= \mathcal{O}_8\phi$ in the counting of lepton number violating operators in \cite{Babu:2001ex}). All tree level completions of the operator $\mathcal{O}_8$ were derived in \cite{Helo:2016vsi} which can be easily adapted to include the SM singlet~$\phi$. These for example include heavy leptoquarks as well as heavy scalars and fermions as present in $R$-parity violating supersymmetry, cf. \suppmat{}~B.2 for more details.
The interactions in Eq.~\eqref{eq:lagrangian} could also be extended in several directions. Most straightforwardly, one can generalize Eq.~\eqref{eq:lagrangian} by including scalar and tensor fermion currents to incorporate all possible Lorentz-invariant combinations. The Majoron may also couple derivatively, if originating as a Goldstone boson; this would increase the number of possible Lorentz-invariant combinations. Alternatively, if the exotic particle is a vector boson $a^\mu$ \cite{Carone:1993jv}, such as a dark photon, the fermion currents can couple to it via the vector field itself as well as its field strength tensor $f^{\mu\nu}$. An even number of exotic neutral fermions $\chi$ may also be emitted but this would quickly increase the dimension of the corresponding effective operator. Instead, they may also originate from the internal neutrino via a dimension-6 operator of the form $\Lambda_\text{NP}^{-2}\nu\nu\chi\chi$~\cite{Huang:2014bva}. Exploring such alternatives to the well-studied neutrinoless double-$\beta$ decay is imperative in order to be able to draw reliable conclusions on the nature of neutrino mass generation.

\begin{acknowledgments}
The work of RC and MH was supported by the Spanish grants
FPA2017-85216-P, SEV-2014-0398 (AEI/FEDER, UE, MICINN), PROMETEO/2018/165 (Generalitat Valenciana), Red
Consolider MultiDark FPA2017-90566-REDC, FPU15/03158 and
EST17/00328. The work of FFD was supported
by the UK STFC and a Royal Society Exchange Grant. The work of LG was
supported by Conicyt Chile under Grant No. 21160645 and DGIIP of
UTFSM. CH acknowledges support within the framework of the
European Union's Horizon 2020 research and innovation programme under
the Marie Sklodowska-Curie grant agreements No 690575 and No 674896. 
RC and LG would like to thank the UCL Department of Physics for
its hospitality.
\end{acknowledgments}

\appendix
\renewcommand{\theequation}{A.\arabic{equation}}
\setcounter{equation}{0}
\section*{A. Calculation of the Exotic Majoron Process}

We here detail the computation of the amplitude and differential decay rate of the $0\nu\beta\beta\phi$ process. We follow the calculation of the standard long-range contributions presented in \cite{Doi:1985dx} and start from the effective Lagrangian
\begin{align}
\label{eq:lagrangianapp}
	\mathcal{L}_{0\nu\beta\beta\phi} = \mathcal{L}_\text{SM} + \mathcal{L}_{R\phi},
\end{align}
with the SM charged current
\begin{align}
	\mathcal{L}_\text{SM} = 
	\frac{G_F\cos\theta_C}{\sqrt{2}}
	j^\mu_L J^{\phantom{\mu}}_{L\mu} + \text{h.c.},
\end{align}
and the exotic 7-dimensional operators incorporating right-handed lepton currents and the Majoron $\phi$,
\begin{align}
	\mathcal{L}_{R\phi} = \frac{G_F\cos\theta_C}{\sqrt{2} m_p}
	\left(
		  \epsilon_{RL}^\phi j_R^\mu J^{\phantom{\mu}}_{L\mu}\phi 
		+ \epsilon_{RR}^\phi j_R^\mu J^{\phantom{\mu}}_{R\mu}\phi  
	\right)
	+ \text{h.c.}.
\end{align}
Here, $G_F$ is the Fermi constant, $\theta_C$ is the Cabbibo angle and the leptonic and hadronic currents are defined as
\begin{align}
	j_{L,R}^\mu = \bar e\gamma^\mu (1\mp\gamma_5) \nu,\quad
	J_{L,R}^\mu = \bar u\gamma^\mu (1\mp\gamma_5) d,
\end{align}
respectively. 

To lowest order of perturbation, the amplitude for the process of $0_I^+ \to 0_F^+$ $0\nu\beta\beta\phi$ decay depicted in Fig.~\ref{fig:diagram}~(center) of the main text is
\begin{align}
	\mathcal{M} = -\int d^4x d^4y \langle F|\mathcal{T} \left\lbrace \mathcal{L}_\text{SM}(x)\mathcal{L}_{R\phi}(y) 
	\right\rbrace |I \rangle.
\end{align}
The time-ordered product is expanded as
\begin{align}
	\mathcal{T} \left\lbrace \mathcal{L}_\text{SM}(x) \mathcal{L}_{R\phi}(y) \right\rbrace 
	&= 2 \, \epsilon_{RX} \frac{(G_F\cos\theta_C)^2}{m_p} \nonumber\\ 
	&\times\mathcal{T} \lbrace J_L^\mu(x) J_X^\nu(y) 
	\underbrace{\bar e(x)\gamma_\mu P_L\nu(x)\bar\nu(y)\gamma_\nu P_L e^c(y)}_{ \Xi^L_{\mu\nu}(x,y)} 
	\phi(y) \rbrace,
\end{align}
with the chiral projectors defined as $P_{L,R} = \frac{1}{2}(1\mp \gamma_5)$. Using the neutrino propagator with momentum $q$ and mass $m_\nu$, the highlighted term $\Xi^L_{\mu\nu}(x,y)$ can be expressed as
\begin{align}
	\Xi^L_{\mu\nu}(x,y) &= 
	\int\frac{d^4 q}{(2\pi)^4} \frac{e^{-iq(x-y)}}{q^2-m_\nu^2 + i\varepsilon} 
	\bar e(x) \gamma_\mu P_L(\slashed{q} + m_\nu)\gamma_\nu P_L e^c(y) \nonumber\\
    &= \int\frac{d^4 q}{(2\pi)^4} q^\alpha 
    \frac{e^{-iq(x-y)}}{q^2 - m_\nu^2 + i\varepsilon} 
    \bar e(x)\gamma_\mu \gamma_\alpha \gamma_\nu P_L e^c(y).
\end{align}
The amplitude needs to be antisymmetric under the exchange of the electrons $e_1$ and $e_2$, and thus we generalize
\begin{align}
	\Xi^{L/R}_{\mu\nu}(x,y) = 
	\frac{1}{\sqrt{2}}\int\frac{d^4 q}{(2\pi)^4} \frac{e^{-iq(x-y)}}{q^2-m_\nu^2+i\varepsilon}
	\left(u^{L/R}_{\mu\nu}(E_1x,E_2y) - u^{L/R}_{\mu\nu}(E_2x,E_1y)\right),
\end{align}
with $u^{L/R}_{\mu\nu}(E_1x, E_2y) = q^\alpha \bar e(E_1,x)\gamma_\mu \gamma_\alpha \gamma_\nu P_{L/R} \, e^c(E_2,y)$ and $E_i$ is the energy of each electron.

We now perform the integral over the temporal variables. The integration over $q_0$ is straightforward by means of the residue theorem,
\begin{align}
	\int\frac{dq_0}{2\pi} \frac{1}{q_0^2-\omega^2} f(q_0) 
	= \frac{i}{2 \omega} f(\omega),
\end{align}
with $\omega^2 = \bf{q}^2 + m_\nu^2$. On the other hand, expanding the time-ordered product as
\begin{equation}
	\mathcal{T} \left\lbrace 
	\mathcal{L}_\text{SM}(x)\mathcal{L}_{R\phi}(y) \right\rbrace 
	= \Theta(x^0-y^0) \mathcal{L}_\text{SM}(x) \mathcal{L}_{R\phi}(y) 
	+ \Theta(y^0-x^0) \mathcal{L}_{R\phi}(y)\mathcal{L}_\text{SM}(x),
\end{equation}
and using the operator $e^{iHt}$ to extract the temporal dependence from the different wave functions, for example $\phi(y) = e^{iE_\phi y_0} \phi(\bf{y})$, one can directly integrate over $x_0$ and $y_0$ obtaining the analogous expression to Eq.~(C.2.19) in \cite{Doi:1985dx},
\begin{align}
\label{eq:M}
	\mathcal{M} = \epsilon_{RX}^{\phi}\frac{(G_F\cos\theta_C)^2}{\sqrt{2}m_p} 
	\sum_N &\int d^3x d^3y \int\frac{d^3q}{2\pi^2\omega} J_{LX}^{\rho\sigma}(\bf{x},\bf{y})\phi(\bf{y}) \nonumber\\
     &\times\left\lbrace\quad\,  e^{i\bf{q}(\bf{x} - \bf{y})}
      \left[ 	
     	  \frac{u_{\rho\sigma}^L(E_1\bf{x},E_2\bf{y})}{\omega+A_2+\frac 12 E_\phi} 
     	- \frac{u_{\sigma\rho}^R(E_1\bf{y},E_2\bf{x})}{\omega+A_1+\frac 12 E_\phi}
      \right]\right. \nonumber\\ 
      &\quad\quad-e^{i\bf{q}(\bf{y}-\bf{x})} 
      \left.\left[
      	  \frac{u_{\rho\sigma}^L(E_1\bf{x},E_2\bf{y})}{\omega+A_1-\frac 12 E_\phi} 
      	- \frac{u_{\sigma\rho}^R(E_1\bf{y},E_2\bf{x})}{\omega+A_2-\frac 12 E_\phi} 
      \right]\right\rbrace,
\end{align}
where $A_{1/2} = E_N - E_I + \frac 12 Q_{\beta\beta} + m_e \pm \frac{1}{2}(E_1 - E_2)$. We anticipate the closure approximation and define the matrix element of the hadronic currents as
\begin{align}
	J_{LX}^{\rho\sigma}(\bf{x},\bf{y}) 
	= \frac 12 \left[ \bra{F} J_L^\rho(\bf{x}) \ket{N}\bra{N} J_X^\sigma(\bf{y}) \ket{I} + \bra{F} J_X^\sigma(\bf{y}) \ket{N}\bra{N} J_L^\rho(\bf{x}) \ket{I} \right].
\end{align}
In addition, the following properties under the exchange of position and electron energies were used in Eq.~\eqref{eq:M},
\begin{equation}
    u^{L/R}_{\rho\sigma}(E_1\bf{x},E_2\bf{y}) = u^{R/L}_{\sigma\rho}(E_1\bf{x},E_2\bf{y}), \qquad J_{LX}^{\rho\sigma}(\bf{x},\bf{y}) = J_{XL}^{\sigma\rho}(\bf{y},\bf{x}).
\end{equation}
The integration over $x_0$ and $y_0$ in Eq.~\eqref{eq:M} also provides the overall energy conservation condition $\delta(Q_{\beta\beta} + 2m_e - E_1 - E_2 - E_\phi)$ with $Q_{\beta\beta} = E_I - E_F - 2m_e$. It is included in the phase space, Eq.~\eqref{eq:dGammaAppendix} below, by requiring $E_\phi = Q_{\beta\beta} + 2m_e - E_1 - E_2$. We additionally assume that the Majoron $\phi$ is emitted predominantly in an $S$-wave configuration, $\phi(\bf{y}) \approx 1$.

Considering the term between braces in Eq.~\eqref{eq:M}, one can write everything under the same exponential by interchanging $\bf{x}$ and $\bf{y}$,
\begin{align}
\label{eq:y}
	e^{i\bf{q}(\bf{x}-\bf{y})} 
	&\left\lbrace \quad\,\left[
	      \frac{J^{\rho\sigma}_{LX}(\bf{x},\bf{y})  
	      u^L_{\rho\sigma}(E_1\bf{x},E_2\bf{y})}{\omega+A_2+\frac{1}{2}E_\phi} 
		+ \frac{J^{\rho\sigma}_{XL}(\bf{x},\bf{y})  
	  	  u^R_{\rho\sigma}(E_1\bf{x},E_2\bf{y})}{\omega+A_2-\frac{1}{2}E_\phi} 
  	\right]\right. \nonumber\\
	&\left.\quad -\left[ 
		  \frac{J^{\rho\sigma}_{LX}(\bf{x},\bf{y})    
	      u^L_{\rho\sigma}(E_2\bf{x},E_1\bf{y})}{\omega+A_1+\frac{1}{2}E_\phi} 
		+ \frac{J^{\rho\sigma}_{XL}(\bf{x},\bf{y}) \; 
		  u^R_{\rho\sigma}(E_2\bf{x},E_1\bf{y})}{\omega+A_1-\frac{1}{2}E_\phi}
	\right]\right\rbrace.
\end{align}
It is furthermore useful to split the leptonic $u^{L,R}_{\rho\sigma}$ functions by separating out the part containing $\gamma_5$ as $u^{L/R}_{\rho\sigma} = \frac{1}{2} \left[u_{\rho\sigma} \mp u^5_{\rho\sigma}\right]$. We then define
\begin{align}
\label{eq:fmunu}
	F^\pm_{\rho\sigma} &= u_{\rho\sigma}(E_1\bf{x},E_2\bf{y}) 
		\pm u_{\sigma\rho}(E_1\bf{y},E_2\bf{x}), \\
    F^{5\pm}_{\rho\sigma} &= u^5_{\rho\sigma}(E_1\bf{x},E_2\bf{y}) 
    	\pm u^5_{\sigma\rho}(E_1\bf{y},E_2\bf{x}), \\
    J^\pm_{\rho\sigma} &= J^{LX}_{\rho\sigma}(E_1\bf{y},E_2\bf{x}) 
    	\pm J^{XL}_{\rho\sigma}(E_1\bf{y},E_2\bf{x}).
    \label{eq:fmunu3}	
\end{align}
These definitions become useful if one recalls that in the non-relativistic impulse approximation, the $J^L$ part of $J^{LX}_{\rho\sigma}$ acts on the $n$-th nucleon whereas the $J^X$ part acts on the $m$-th when performing the sum over all neutrons in the initial nucleus. The superscript $\pm$ in $J^\pm_{\rho\sigma}$ thus indicates if the combination of currents is symmetric or antisymmetric under the interchange of $m\leftrightarrow n$. The same applies to $F^\pm_{\rho\sigma}$ and $F^{5\pm}_{\rho\sigma}$.

The closure approximation implies that the sum over all possible intermediate states is performed analytically using the completeness of all intermediate states and by replacing the intermediate state energies $E_N$ with a common average $\langle E_N \rangle$. This means that the antisymmetric combinations under the interchange of the nucleons $m$ and $n$ will vanish, as the sum is performed over all possible configurations. From Eqs.~\eqref{eq:M} and \eqref{eq:y}, the non-vanishing terms are
\begin{align}
	\mathcal{M} &= \epsilon_{RX} \frac{(G_F\cos\theta_C)^2}{2\sqrt{2}m_p} 
	 \nonumber\\
	&\times\sum_N (H_{\omega 2} - H_{\omega 1})\left\lbrace 
		  J_{\mu\nu}^+ F^{+,\mu\nu}   
		- J_{\mu\nu}^- F^{5-,\mu\nu}
		+ \frac{E_\phi}{E_{12}}
		  \left(J_{\mu\nu}^+ F^{5+,\mu\nu} - J_{\mu\nu}^-F^{-,\mu\nu}
		  \right)\right\rbrace,
\end{align}
where $H_{\omega i}$ are neutrino potentials defined as
\begin{align} \label{eq:Hw}
	H_{\omega i} = \int \frac{d^3q}{2\pi^2\omega} \frac{\omega}{\omega + A_i} e^{i\bf{q}(\bf{x}-\bf{y})}.
\end{align}
Now, the connection with the results of \cite{Doi:1985dx} can be done by contracting the leptonic and nuclear currents within the impulse approximation. The only change in our case is in the $\omega$ term,
\begin{align} 
\label{eq:M2}
    \mathcal{M}_\omega &\propto (H_{\omega 2} - H_{\omega 1})
    \left\lbrace 
    	  (X_3 + X_{5 R})\left[F_+^0 + \frac{E_{\phi}}{E_{12}}F_{5 +}^0\right] 
    	+ Y_{3R}\left[F_{5 -}^0 + \frac{E_{\phi}}{E_{12}}F_{-}^0\right]\right. 
    \nonumber\\
 	&+ \left.(X_{4 R}^l + X_{5}^l)\left[F_+^l +\frac{E_{\phi}}{E_{12}}F_{5 +}^l \right] 
 	 + (Y_4^l -Y_{5R}^l)\left[F_{5-}^l 
 	 + \frac{E_{\phi}}{E_{12}}F_{-}^l\right]\right\rbrace,
\end{align} 
where the $X$ and $Y$ terms are functions of nuclear parameters and operators defined in Appendix~C of \cite{Doi:1985dx}. The $F^{\alpha}_{(5)\pm}$-terms are generated by the contraction of the hadronic and leptonic parts in Eqs.~\eqref{eq:fmunu}-\eqref{eq:fmunu3} factorizing out the dependence with the momentum $q_{\alpha}$ from the leptonic part (see  Eq.~(C.2.25) in \cite{Doi:1985dx}). One trivially recovers the $\omega$ term in the expression~(C.2.23) of \cite{Doi:1985dx} for $E_\phi \to 0$.

Comparing Eq.~\eqref{eq:M2} with the results from \cite{Doi:1985dx}, one can track the dependence with $E_\phi$ in the decay rate down to Eq.~(C.3.9) of \cite{Doi:1985dx}. The main change for $0^+_I\to 0^+_F$ transitions is in the terms $N_3$ and $N_4$ where a contribution proportional to $E_{\phi}$ appears explicitly, 
\begin{align}
\label{eq:N12}
	\begin{pmatrix}
		N_1 \\
		N_2
	\end{pmatrix}
	&=
	\begin{pmatrix}
		\alpha_{-1-1}^* \\
		\alpha_{11}^*
	\end{pmatrix}
	\left[ \frac{4}{3}Z_6
	\mp \frac{4}{m_e R}\left(Z_{4R}-\frac{1}{6} \zeta Z_6\right)\right], \\
	\begin{pmatrix}
		N_3 \\
		N_4
	\end{pmatrix}
	&=
	\begin{pmatrix}
		\alpha_{1-1}^* \\
		\alpha_{-11}^*
	\end{pmatrix}
	\left[ -\frac{2}{3}Z_5 
	\mp \frac{E_{12}}{m_e}
	\left(Z_3 + \frac{1}{3}Z_5\right)+\frac{E_{\phi}}{m_e}Z_3 \right].
\label{eq:N34}
\end{align}
Here, $\alpha_{jk} = \tilde{A}_j(E_1) \tilde{A}_k(E_2)$ describe the Coulomb-corrected relativistic electron wave functions and $\zeta = 3\alpha Z + (Q_{\beta\beta} + 2m_e)R$ the correction of the electron $P$ wave, with the fine structure constant $\alpha$ and the radius $R$ and charge $Z$ of the final state nucleus. The information about the electron wave functions is encoded in 
\begin{align}
	\tilde{A}_{\pm k}(E) = \sqrt{\frac{E \mp m_e}{2E} F_{k-1}(Z,E)},
\end{align}
with the Fermi factor
\begin{align}
	F_{k-1}(Z,E) = \left[\frac{\Gamma(2k+1)}{\Gamma(k)\Gamma(2\gamma_k+1)}\right]^2(2pR)^{2(\gamma_k -k)}|\Gamma(\gamma_k+ iy)|^2e^{\pi y},
\end{align}
where $\gamma_k = \sqrt{k^2 + (\alpha Z)^2}$, $y = \alpha Z E/p$ and $p = \sqrt{E^2 - m_e^2}$.

In order to arrive at Eq.~\eqref{eq:M2} one should neglect the higher order terms $E_{12}^2$, $E_{12}E_\phi$ and $E_\phi^2$ as they are suppressed with an extra denominator $(\omega+A_i)$ compared to Eq.~\eqref{eq:Hw}.

\setlength{\tabcolsep}{5pt}
\begin{table}[t!]
	\centering
	\begin{tabular}{ccccccccccc}
		\hline
		Isotope & $Q_{\beta\beta}$ [MeV] & $M_{GT}$ & $\chi_F$ & $\chi_{GT\omega}$ & $\chi_{F\omega}$ & $\chi_{GT}'$ & $\chi_F'$ & $\chi_T$ & $\chi_R$ & $\chi_P$ \\
		\hline
		$^{82}\text{Se}$ & 2.99 & 2.993 & $-0.134$ & 0.947 & $-0.131$ & 1.003 & $-0.103$ & $\phantom{-}0.004$ & 1.086 & 0.430 \\
		$^{136}\text{Xe}$ & 2.46 & 1.770 & $-0.158$ & 0.908 & $-0.149$ & 1.092 & $-0.167$ & $-0.031$ & 0.955 & 0.256 \\
		\hline
	\end{tabular}
	\caption{Energy release $Q_{\beta\beta}$ and relevant nuclear matrix elements for $^{82}\text{Se}$ and $^{136}\text{Xe}$ used in the calculation of the $0\nu\beta\beta\phi$ decay rate and distributions. The nuclear matrix elements were taken from the shell model calculations \cite{Horoi:2015gdv} ($^{82}$Se) and \cite{Caurier:1996bu} ($^{136}$Xe), except for $M_{GT}$ in $^{136}\text{Xe}$ where we use an updated value from the same group \cite{Caurier:2007qn}.}
	\label{table:NMEs}
\end{table}
The $Z_i$ terms are given in Eqs.~\eqref{eq:Zs}-\eqref{eq:Zs4} below and they contain the nuclear matrix elements and effective particle physics couplings. The $Z_i$ terms are the same as in \cite{Doi:1985dx}, with the relevant couplings $\lambda \to \epsilon_{RR}^{\phi}$ and $\eta \to \epsilon_{RL}^{\phi}$ substituted. Note that the term with $Z_1$ in Eq.~(C.3.9) from \cite{Doi:1985dx} related to the standard $0\nu\beta\beta$ decay disappears from Eq.~\eqref{eq:N12}, as we are not considering the interaction $\mathcal{L}_{SM}(x) \mathcal{L}_{SM}(y)$.
\begin{align} 
\label{eq:Zs}
    Z_3 &= \left[-\epsilon_{RR}^{\phi}(\chi_{GT\omega}-\chi_{F\omega}) + \epsilon_{RL}^{\phi}(\chi_{GT\omega}+\chi_{F\omega}) \right] M_{GT},
    \\
    Z_{4R} &= \epsilon_{RL}^{\phi} \chi_R M_{GT},
    \\
    Z_5 &= \frac 13 \left[ \epsilon_{RR}^{\phi}(\chi'_{GT}-6\chi_T+3\chi'_F) - \epsilon_{RL}^{\phi}(\chi'_{GT}-6\chi_T-3\chi'_F) \right] M_{GT},
    \\
    Z_6 &= \epsilon_{RL}^{\phi} \chi_P M_{GT}.
   \label{eq:Zs4}
\end{align}
The above equations are valid when both $\epsilon^\phi_{RL}$ and $\epsilon^\phi_{RR}$ are present. For our numerical calculations, we use the $Q_{\beta\beta}$ values and nuclear matrix elements $M_{GT}$, $\chi_F$, etc. presented in Table~\ref{table:NMEs} for $^{82}\text{Se}$ and $^{136}\text{Xe}$. We use the following values for the remaining parameters: $G_F = 1.2 \times 10^{-5}~\text{GeV}^{-2}$, $\alpha = 1/137$, $g_A = 1.27$, $R = 1.2 A^{1/3}$~fm with the mass number $A$ of the isotope in question. The factors $N_1$, $N_2$, $N_3$ and $N_4$ in Eqs.~\eqref{eq:N12} and \eqref{eq:N34} are then fully described and the energy-dependent coefficients are
\begin{align}
	a(E_1, E_2, E_\phi) &= |N_1|^2 + |N_2|^2 + |N_3|^2 + |N_4|^2,\\
	b(E_1, E_2, E_\phi) &= -2\,\text{Re}\left(N_1^* N_2 + N_3^* N_4\right).
\end{align}
The differential decay rate for the $0^+\to 0^+$ $0\nu\beta\beta\phi$ decay can then be written as \cite{Doi:1985dx}
\begin{align}
\label{eq:dGammaAppendix}
	d\Gamma = C
	\left[a(E_1,E_2,E_\phi) + b(E_1,E_2,E_\phi) \cos\theta\right] 
	w(E_1, E_2, E_\phi) \, dE_1 \, dE_2 \, d\!\cos\theta ,
\end{align}
with 
\begin{gather}
	C = \frac{(G_F\cos\theta_C g_A)^4 m_e^{9}}{256\pi^7 (m_p R)^2}, \\
	w(E_1, E_2, E_\phi) = m_e^{-7} p_1 p_2 E_1 E_2 E_\phi.
\end{gather}
Here, $g_A$ is the axial coupling of the nucleon and $R$ is the radius of the nucleus. The magnitudes of the electron momenta are given by $p_i = \sqrt{E_i^2 - m_e^2}$ and $0 \leq \theta \leq \pi$ is the angle between the emitted electrons. Throughout the above expressions, the Majoron energy is implicitly fixed by the electron energies as $E_\phi = Q_{\beta\beta} + 2m_e - E_1 - E_2$ due to overall energy conservation.

The total decay rate $\Gamma$ and the half life $T_{1/2}$ are then calculated as
\begin{align}
\label{eq:TotalRate}
	\Gamma = \frac{\ln 2}{T_{1/2}} = 2 C
	\int_{m_e}^{Q_{\beta\beta} + m_e} dE_1 \int_{m_e}^{Q_{\beta\beta} + 2m_e - E_1} dE_2 \, 
	a(E_1, E_2, E_\phi) w(E_1, E_2, E_\phi).
\end{align}
\begin{figure}[t!]
	\centering
	\includegraphics[clip,trim={70 20 50 20},width=\textwidth]{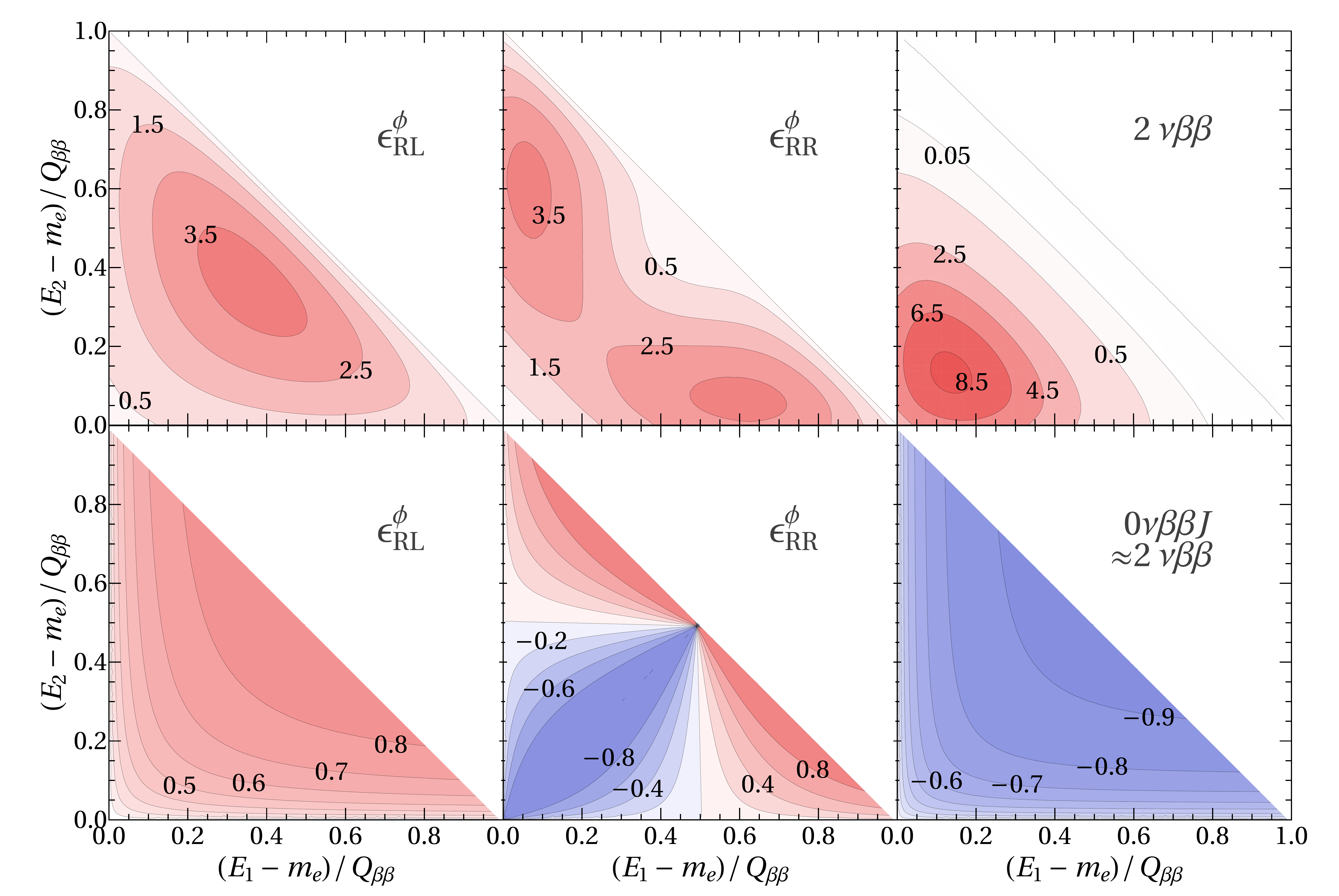}
	\caption{Double electron energy distribution $\frac{d\Gamma}{dE_1dE_2}$ (top row) and electron angular correlation $\alpha$ (bottom row) as function of the individual electron kinetic energies for $^{82}\text{Se}$. Each column is for a specific scenario: $0\nu\beta\beta\phi$ Majoron emission through $\epsilon^\phi_{RL}$ (left) and $\epsilon^\phi_{RR}$ (center); SM $2\nu\beta\beta$ decay (right). The angular correlation of the latter is approximately identical to ordinary Majoron emission $0\nu\beta\beta J$.}
	\label{fig:angular2}
\end{figure}
The fully differential energy information is encoded in the normalized double energy distribution
\begin{align}
	\Gamma^{-1}\frac{d\Gamma}{dE_1 dE_2} 
	= \frac{2 C}{\Gamma} a(E_1, E_2, E_\phi) w(E_1, E_2, E_\phi).
\end{align}
This function, in terms of the kinetic energies normalized to the $Q$ value, $(E_i - m_e)/Q_{\beta\beta}$, is plotted in the top row of Fig.~\ref{fig:angular2} for the case of $0\nu\beta\beta\phi$ Majoron emission through $\epsilon^\phi_{RL}$ (left) and $\epsilon^\phi_{RR}$ (center) as well as for the SM $2\nu\beta\beta$ decay (right). The plots are for the isotope $^{82}\text{Se}$ but would be qualitatively similar for $^{136}\text{Xe}$. As can be seen, the shapes depicted as contours are different between all three modes. Especially the $\epsilon^\phi_{RR}$ exhibits an asymmetry in that one of the electrons takes the majority of the visible energy. If the individual electron energies can be measured, as e.g. in the NEMO-3 or SuperNEMO experiments, this can be exploited to enhance the signal over the $2\nu\beta\beta$ background. As an illustrating example, requiring that any one of the electrons in a signal event has a kinetic energy $E_i - m_e > Q_{\beta\beta} / 2$ would reduce the $0\nu\beta\beta\phi$-$\epsilon^\phi_{RR}$ rate only by a factor of 2 but would suppress the $2\nu\beta\beta$ rate by a factor of 20. The distributions in Fig.~\ref{fig:energydist} of the main text can be easily determined by appropriately integrating over $\frac{d\Gamma}{dE_1 dE_2}$.

In addition to the energies, the angle between the electron momenta also contains useful information. The so-called angular correlation defined by 
\begin{align}
	\alpha(E_1, E_2) 
	= \frac{b(E_1, E_2, E_\phi)}{a(E_1, E_2, E_\phi)},
\end{align}
is a function of the individual electron energies which can take values between $-1$ (the two electrons are dominantly emitted back-to-back) and $+1$ (the two electrons are dominantly emitted collinearly). For $^{82}\text{Se}$ it is plotted in the bottom row of Fig.~\ref{fig:angular2} in the three modes of interest. As expected from angular momentum considerations, the electrons are dominantly emitted back-to-back in the SM $2\nu\beta\beta$ decay with $(V-A)$ lepton currents, $\alpha < 0$ for all energies. For $\epsilon^\phi_{RL}$, they are dominantly emitted collinearly, $\alpha > 0$ for all energies. In the case of $\epsilon^\phi_{RR}$, the behaviour is complex due to the asymmetry of the amplitude under the exchange of electrons and nuclear recoil effects. The correlation $\alpha$ changes sign, with $\alpha > 0$ when any one electron has a kinetic energy $E_i - m_e > Q_{\beta\beta} / 2$ but $\alpha <0$ when both electrons each have a kinetic energy $E_i - m_e < Q_{\beta\beta} / 2$. Note that Fig.~\ref{fig:angular2} provides the full kinematical information in each mode; all measurable quantities can be constructed from these distributions.

As discussed in the main text, averaged over all energies, the electrons are actually emitted almost isotropically for $\epsilon^\phi_{RR}$. This is quantified by integrating Eq.~\eqref{eq:dGammaAppendix} over all energies analogous to Eq.~\eqref{eq:TotalRate} to yield the angular distribution
\begin{align}
	\frac{d\Gamma}{d\cos\theta} = \frac{\Gamma}{2}(1 + k\,\cos\theta),
\end{align}
with the average angular correlation factor $k$.

\section*{B. Ultraviolet Complete Scenarios}

\subsection{Left-right symmetry}
Here we briefly discuss the Left-Right symmetric scenario mentioned in the main text as a possible ultraviolet completion generating the effective operators in Eq.~(1) in the main text. At some energy scale above the electroweak scale we assume a left-right symmetric gauge symmetry
\begin{align}
	G_\text{LR} \equiv SU(3)_c \times SU(2)_L \times SU(2)_R \times U(1)_X.
\end{align}
It breaks down to the SM gauge group $G_\text{SM} \equiv SU(3)_c \times SU(2)_L \times U(1)_Y$ at a scale $M_R$. The SM electric charge is related to the generators of the gauge groups by the relation
\begin{align}
\label{1.1}
	Q = T_{3L} + T_{3R} + \frac{X}{2} = T_{3L} + Y.
\end{align}
In the minimal Left-Right symmetric model, the quantum number $X$ is identified with $B-L$, i.e. $B-L$ is a gauge symmetry in the model. Consequently, left-right symmetry breaking is usually assumed to induce several $B-L$ violating interactions, including generation of Majorana neutrino masses via a seesaw mechanism. 

In general, though, one can define a new quantum number $\zeta$ such that
\begin{align}
\label{1.2}
	X = (B-L) + \zeta.
\end{align}
If $\zeta \neq 0$ then $B-L$ can remain a global symmetry, independent of the left-right gauge symmetry. We here propose such a scenario where the field content and their quantum numbers for such a realisation of a Left-Right symmetric model are summarised in Table~\ref{tab:LR2}.
\begin{table}[t!]
\centering
	\begin{tabular}{c|cccccc}
		\hline
		Field     & $SU(2)_L$ & $SU(2)_R$ & $B-L$ & $\zeta$ & $X$ & $SU(3)_C$ \\
		\hline
		$q_L$     &  2 & 1 & 1/3 &  0 &  1/3 & 3 \\
		$q_R$     &  1 & 2 & 1/3 &  0 &  1/3 & 3 \\
		$\ell_L$  &  2 & 1 &  -1 &  0 &   -1 & 1 \\
		$\ell_R$  &  1 & 2 &  -1 &  0 &   -1 & 1 \\
		\hline
		$U_{L,R}$ &  1 & 1 & 1/3 & +1 &  4/3 & 3 \\
		$D_{L,R}$ &  1 & 1 & 1/3 & -1 & -2/3 & 3 \\
		$E_{L,R}$ &  1 & 1 &  -1 & -1 &   -2 & 1 \\
		$N_{L,R}$ &  1 & 1 &  -1 & +1 &    0 & 1 \\
		\hline
		$\chi_L$  &  2 & 1 &   0 & +1 &    1 & 1 \\
		$\chi_R$  &  1 & 2 &   0 & +1 &    1 & 1 \\
		\hline
		$\phi$    & 1  & 1 &   2 & -2 &    0 & 1 \\
		\hline
	\end{tabular}
	\caption{Field content and quantum numbers under $G_\text{LR}$ in the Left-Right symmetric scenario proposed.}
	\label{tab:LR2}
\end{table}
Apart from the SM fields $q_L$, $\ell_L$ and the SM quark singlets which transform as a doublet $q_R$ under $G_\text{LR}$, left-right symmetry naturally includes right-handed neutrino fields $\nu_R$ as part of the right handed lepton doublet $\ell_R$. 

As noted, the assumed symmetry breaking pattern is given by
\begin{align}
\label{1.4}
	G_\text{LR} \stackrel{M_R}{\to} G_\text{SM}
	\stackrel{m_W}{\to} SU(3)_c \times U(1)_Q.
\end{align}
For the left-right symmetry breaking, we use a doublet Higgs scalars $\chi_R$, whose vacuum expectation value (VEV) breaks the left-right symmetry \cite{Senjanovic:1978ev, Dvali:1997uq, Babu:1998wi, Barr:2003nn, Albright:2003xb, Albright:2004ws}. This field does not have any exclusive interaction with the SM fermions and hence the $B-L$ quantum number is no longer uniquely determined. Thus for $\chi_R$, we can choose $B-L = 0$, and hence, $\zeta = 1$ in Eq.~\eqref{1.2}. Note that our model differs from earlier models in this choice of the $B-L$ quantum number \footnote{Nevertheless, the model remains anomaly free under the charge $X$ as the new vector-like fermions do not contribute to the anomalies and the assignments of $X$ for the chiral fermions are the same as that of $B-L$ in the conventional Left-Right symmetric model.}. The left-right symmetry ensures that we have a second doublet Higgs scalar $\chi_L$ with the same assignment of $B-L = 0$ and $\zeta = 1$. Interestingly, these assignments do not require any additional global symmetries, but will allow $B-L$ to remain as a global symmetry after the electroweak symmetry breaking.

A priori, we have two choices of Higgs scalar for breaking the electroweak symmetry. The first choice is that we retain the Higgs bi-doublet from the conventional model, which, after electroweak symmetry breaking, will generate Dirac masses for all the fermions. One particularly interesting scenario arises if we assume that only quarks acquire their masses through the VEV of the bi-doublet and the Yukawa couplings giving rise to such masses of leptons are forbidden by some symmetry\footnote{For example one may introduce an additional discrete $Z_2$ symmetry such that $\ell_R$, $N_R$ and $E_R$ are odd under this discrete symmetry. Note that in such a case the vector-like mass term for $N$ and $E$ (see. Eq.~\eqref{lag:dss}) will break this $Z_2$ symmetry softly.}. Both the charged and the neutral leptons would then acquire Dirac seesaw masses in this scenario \cite{Davidson:1987mh, Rajpoot:1987ji, Chang:1986bp, Balakrishna:1987qd, Babu:1988mw, Babu:1989rb, Brahmachari:2003wv}. An alternative is that there is no Higgs bi-doublet and the left-handed Higgs doublet $\chi_L$ breaks the electroweak symmetry. In such a scenario the quark masses and the charged lepton masses are generated through a seesaw mechanism introducing new vector-like states. This scheme is often called the universal seesaw mechanism. We will mainly focus on this second scenario, however all the discussion presented are applicable for both the scenarios.

In the leptonic sector, we introduce four singlet vector-like fermions, which are the charged and neutral heavy leptons $N_L$, $N_R$, $E_L$, $E_R$ in Table~\ref{tab:LR2},
all carrying $B-L = 1$, and hence $\zeta = -1$ for the neutral fermions $N_{L,R}$ and $\zeta = 1$ for the charged fermions $E_{L,R}$. The left-right symmetry breaking will allow mixing of these fermions with the light leptons and the assignment of lepton number is somewhat more natural than in conventional left-right symmetric models where similar new singlets carry vanishing lepton numbers. The VEVs $u_{L,R}$ of the fields $\chi_{L,R}$ introduce mixing of the new neutral leptons $\sigma_{L,R}$ with the neutrinos and the new charged leptons $E_{L,R}$ with the charged leptons.

In the absence of the Higgs bi-doublet, $\chi_L$ breaks the electroweak symmetry. In this case we need to introduce vector-like states for all fermions in order to generate their masses. Consequently, all the masses of quarks, leptons and neutrinos are generated by a Dirac seesaw mechanism. 

For the charged and neutral leptons there are no bare Dirac mass terms. The Yukawa interactions that give masses to the leptons are given by
\begin{align}
\label{lag:dss}
	{\cal L}_Y &= 
	   \ell_L^T \cdot f_L \cdot C^{-1} N_L \;\chi_L 
	 + \ell_R^T \cdot f_R \cdot C^{-1} N_R \;\chi_R 
	 + \bar N_L \cdot m_N \cdot N_R            \nonumber\\
	&+ \bar\ell_L \cdot h_L \cdot E_R \;\chi_L
	 + \bar\ell_R \cdot h_R \cdot E_L \;\chi_R
	 + \bar E_L \cdot m_E \cdot E_R + \text{h.c.}.
\end{align}
Here, we suppress generation indices and thus $f_{L,R}$, $m_N$, $h_{L,R}$ and $m_E$ are $3\times 3$ matrices in the space of fermion generations. The charged lepton masses are generated through a Dirac seesaw mechanism and the mass matrix is given by
\begin{align}
\label{lepton:mass}
	m_\ell = u_L u_R \left(h_L \cdot m^{-1}_E \cdot h^\dagger_R \right).
\end{align}
 
In discussing the fields associated with neutrinos, we now work with the CP conjugates of the right-handed fields for convenience,
\begin{align}
	\nu_R \stackrel{CP}{\to} (\nu_R)^c = (\nu^c)_L = \sigma_L, \quad 
	  N_R \stackrel{CP}{\to} (N^c)_L = \Sigma_L,
\end{align}
so that the mass matrix of the fields $(\nu_L, \sigma_L, N_L, \Sigma_L)^T$ can be written as 
\begin{equation}
	{\cal M}_\nu = 
	\begin{pmatrix}
		0        & 0       & u_L f_L & 0       \\ 
		0        & 0       & 0       & u_R f_R \\
		u_L f_L  & 0       & 0       & m_N     \\ 
		0        & u_R f_R & m_N     & 0       \\
	\end{pmatrix}.
\end{equation}
This results in a spectrum of six Dirac neutrinos, three very heavy with masses $\approx m_N$ and three light with masses $\approx u_L u_R f_L f_R/m_N$. The heavy Dirac neutrinos are composed of $N_L$ and $\Sigma_L$ while the light Dirac neutrinos represent the active SM neutrinos in a combination of $\nu_L$ and $\sigma_L$ or $\nu_R$. Unlike the usual models of light Dirac neutrinos \cite{Murayama:2002je, Thomas:2005rs, Thomas:2006gr, Abel:2006hr, Cerdeno:2006ha, Gu:2006dc, Gu:2007mc, Gu:2007mi}, where the light neutrino masses are proportional to a small induced VEV, in this scenario the smallness of neutrino masses is due to a seesaw mechanism where the heavy seesaw scale corresponds to the masses of the vector-like neutrino states. 

Finally, the last ingredient of this model is a light charge-neutral scalar particle $\phi$ that can potentially be a Dark Matter candidate \cite{Berezinsky:1993fm, Garcia-Cely:2017oco, Brune:2018sab}. Of main interest to us, the presence of such a particle with a Yukawa coupling to $N$ of the form $g_\phi N N \phi$, can lead to $0\nu\beta\beta\phi$ decay with emission of a light neutral scalar $\phi$ from a single effective dimension-7 operator of the form $\Lambda_\text{NP}^{-3}(\bar u\mathcal{O} d)(\bar e\mathcal{O}\nu)\phi$ as discussed in the main text. This model provides a working example of a scenario where purely Dirac neutrinos can mimic conventional $0\nu\beta\beta$ decay through emission of extra particle that carries lepton number. This illustrates the necessity of searches for extra particles in double-$\beta$ decay to understand the nature of neutrinos.

\subsection{Leptoquarks and ${\mathbf R}$-parity violating supersymmetry}
Here we briefly discuss an alternative scenario for generating the effective Majoron current in Eq.~(1) of the main text. This setup is based on a simple extension of the SM, introducing two heavy scalar leptoquarks $S_{3,2,1/6}$, $S_{3^*,1,1/3}$ and a scalar singlet $\phi$. Here, the quantum numbers of the leptoquarks under the SM gauge group $SU(3)_C \times SU(2)_L \times U(1)_Y$ are as indicated. The interesting part of the Lagrangian is given by
\begin{equation}
\label{eq:LagBL8}
	{\cal L}^\text{LQ} = 
	  Y^1_{\alpha\beta} L_\alpha \bar d_{R\beta} S_{3,2,1/6}
	+ Y^2_{\alpha\beta} e_{R\alpha} u_{R\beta} S_{3^*,1,1/3}
	+ Y^S \phi H S_{3,2,1/6}^\dagger S_{3^*,1,1/3}^\dagger.
\end{equation}
Assigning lepton numbers as $L(S_{3,2,1/6}) = -1$, $L(S_{3^*,1,1/3}) = -1$ and $L(\phi) = -2$ this Lagrangian \emph{conserves} lepton number. In Eq.~\eqref{eq:LagBL8}, we have written out explicitly generation indices of the lepton and quark fields, $\alpha,\beta=1,2,3$. Thus, $Y^1$ and $Y^2$ are in general $3\times 3$ matrices. However, double-$\beta$ decay will be sensitive only to first generation.

\begin{figure}[t!]
\centering
\includegraphics[width=0.45\textwidth]{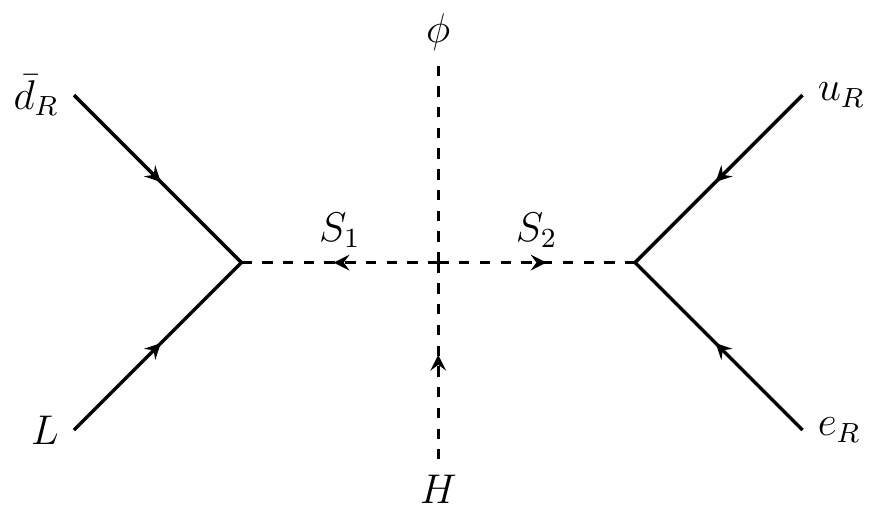}
\includegraphics[width=0.5\textwidth]{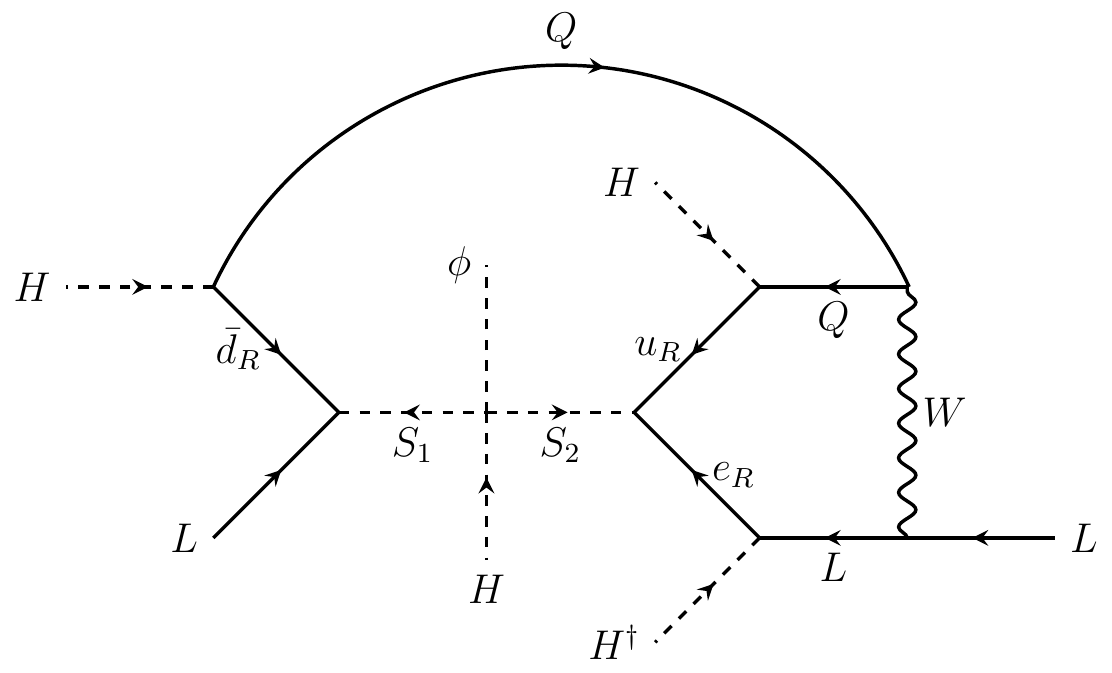}
\caption{Possible leptoquark contribution to the decay $\znbb\phi$ (left). If $\phi$ develops a VEV, the 2-loop diagram (right) is unavoidable in this model. Note that the presence of this singlet VEV signals lepton number violation.}
\label{fig:BL8}
\end{figure}

Integrating out the heavy leptoquark states, this Lagrangian leads to an effective coupling $\epsilon_{\alpha\beta} ({\bar d_R}L^{\alpha}) ({\bar e_R^c}u_R) H^{\beta} \phi$. After electroweak symmetry breaking and Fierz rearrangement of the fields at low energies, the effective current $\frac{\epsilon^\phi_{RR}}{m_p} j_R^\mu J^{\phantom{\mu}}_{R\mu}\phi$ is generated as shown in Fig.~\ref{fig:BL8}~(right). 

The phenomenology of this setup depends on whether the scalar $\phi$ develops a VEV,
\begin{align}
\label{eq:LQsp} 
	Y^S \Phi H S_{3,2,1/6}^\dagger S_{3^*,1,1/3}^\dagger
	\Rightarrow 
	Y^S \langle\phi\rangle H S_{3,2,1/6}^\dagger S_{3^*,1,1/3}^\dagger.
\end{align}
In this case, lepton number is spontaneously broken and a massless (exotic) Majoron appears automatically and Majorana neutrino masses are generated. In Fig.~\ref{fig:BL8}~(right) we show the 2-loop neutrino mass diagram, which will 
result unavoidably for $\langle \phi \rangle \ne 0$. A rough estimate of the neutrino mass generated by this diagram is
\begin{align}
\label{eq:mnu2lp}
	m_\nu \approx 
	\frac{Y^1 Y^2 m_u m_d m_\ell}{(16\pi^2)^2} 
	\frac{Y^S \langle\phi\rangle v_{SM}}{\Lambda^4},
\end{align}
where $\Lambda$ is of the order of the leptoquark masses and $m_u$, $m_d$, $m_\ell$ indicate the SM quark and lepton masses. For couplings of order one, 3rd generation SM fermion masses and $\Lambda = {\cal O}(1)$~TeV, neutrino masses of the order of the atmospheric scale can be generated for $\langle\phi\rangle \approx 10$~GeV. However, due to the smallness of the first generation fermion masses, no constraint on their couplings to leptoquarks can be derived from neutrino masses.

As in all such Majoron models, whether the constraints from non-observation of $\znbb\phi$ are more important than neutrino mass constraints or from the non-observation of ordinary $\znbb$ depends on the unknown value of $\langle \phi \rangle$. For $\langle \phi \rangle$ approaching zero lepton number is effectively restored and at low energies $\znbb\phi$ will provide the only constraint.

Finally, we would like to remark that the two leptoquarks in this model have the same quantum numbers as the scalar quark doublet and down-type scalar quark singlet fields, $S_{3,2,1/6} \equiv \tilde Q$ and $S_{3^*,1,1/3} \equiv \tilde d^c$ in supersymmetric models. This opens up to possibility to speculate about Majoron model variants in $R$-parity violating supersymmetry. However, different from the model discussed above, for $R$-parity violating supersymmetry the Lagrangian would contain terms $L_\alpha Q_\beta S_{3^*,1,1/3}$ instead of $e_{R\alpha} u_{R\beta} S_{3^*,1,1/3}$. This affects the discussion of the phenomenology, since (i) neutrino masses become 1-loop effects and (ii) at low energies different currents from the ones considered in the main text would be generated.

\bibliography{references}

\begin{thebibliography}{50}%
\makeatletter
\providecommand \@ifxundefined [1]{%
 \@ifx{#1\undefined}
}%
\providecommand \@ifnum [1]{%
 \ifnum #1\expandafter \@firstoftwo
 \else \expandafter \@secondoftwo
 \fi
}%
\providecommand \@ifx [1]{%
 \ifx #1\expandafter \@firstoftwo
 \else \expandafter \@secondoftwo
 \fi
}%
\providecommand \natexlab [1]{#1}%
\providecommand \enquote  [1]{``#1''}%
\providecommand \bibnamefont  [1]{#1}%
\providecommand \bibfnamefont [1]{#1}%
\providecommand \citenamefont [1]{#1}%
\providecommand \href@noop [0]{\@secondoftwo}%
\providecommand \href [0]{\begingroup \@sanitize@url \@href}%
\providecommand \@href[1]{\@@startlink{#1}\@@href}%
\providecommand \@@href[1]{\endgroup#1\@@endlink}%
\providecommand \@sanitize@url [0]{\catcode `\\12\catcode `\$12\catcode
  `\&12\catcode `\#12\catcode `\^12\catcode `\_12\catcode `\%12\relax}%
\providecommand \@@startlink[1]{}%
\providecommand \@@endlink[0]{}%
\providecommand \url  [0]{\begingroup\@sanitize@url \@url }%
\providecommand \@url [1]{\endgroup\@href {#1}{\urlprefix }}%
\providecommand \urlprefix  [0]{URL }%
\providecommand \Eprint [0]{\href }%
\providecommand \doibase [0]{http://dx.doi.org/}%
\providecommand \selectlanguage [0]{\@gobble}%
\providecommand \bibinfo  [0]{\@secondoftwo}%
\providecommand \bibfield  [0]{\@secondoftwo}%
\providecommand \translation [1]{[#1]}%
\providecommand \BibitemOpen [0]{}%
\providecommand \bibitemStop [0]{}%
\providecommand \bibitemNoStop [0]{.\EOS\space}%
\providecommand \EOS [0]{\spacefactor3000\relax}%
\providecommand \BibitemShut  [1]{\csname bibitem#1\endcsname}%
\let\auto@bib@innerbib\@empty
\bibitem [{\citenamefont {Chikashige}\ \emph {et~al.}(1981)\citenamefont
  {Chikashige}, \citenamefont {Mohapatra},\ and\ \citenamefont
  {Peccei}}]{Chikashige:1980ui}%
  \BibitemOpen
  \bibfield  {author} {\bibinfo {author} {\bibfnamefont {Y.}~\bibnamefont
  {Chikashige}}, \bibinfo {author} {\bibfnamefont {R.~N.}\ \bibnamefont
  {Mohapatra}}, \ and\ \bibinfo {author} {\bibfnamefont {R.~D.}\ \bibnamefont
  {Peccei}},\ }\href {\doibase 10.1016/0370-2693(81)90011-3} {\bibfield
  {journal} {\bibinfo  {journal} {Phys. Lett.}\ }\textbf {\bibinfo {volume}
  {98B}},\ \bibinfo {pages} {265} (\bibinfo {year} {1981})}\BibitemShut
  {NoStop}%
\bibitem [{\citenamefont {Gelmini}\ and\ \citenamefont
  {Roncadelli}(1981)}]{Gelmini:1980re}%
  \BibitemOpen
  \bibfield  {author} {\bibinfo {author} {\bibfnamefont {G.~B.}\ \bibnamefont
  {Gelmini}}\ and\ \bibinfo {author} {\bibfnamefont {M.}~\bibnamefont
  {Roncadelli}},\ }\href {\doibase 10.1016/0370-2693(81)90559-1} {\bibfield
  {journal} {\bibinfo  {journal} {Phys. Lett.}\ }\textbf {\bibinfo {volume}
  {99B}},\ \bibinfo {pages} {411} (\bibinfo {year} {1981})}\BibitemShut
  {NoStop}%
\bibitem [{\citenamefont {Carone}(1993)}]{Carone:1993jv}%
  \BibitemOpen
  \bibfield  {author} {\bibinfo {author} {\bibfnamefont {C.~D.}\ \bibnamefont
  {Carone}},\ }\href {\doibase 10.1016/0370-2693(93)90605-H} {\bibfield
  {journal} {\bibinfo  {journal} {Phys. Lett.}\ }\textbf {\bibinfo {volume}
  {B308}},\ \bibinfo {pages} {85} (\bibinfo {year} {1993})},\ \Eprint
  {http://arxiv.org/abs/hep-ph/9302290} {arXiv:hep-ph/9302290 [hep-ph]}
  \BibitemShut {NoStop}%
\bibitem [{\citenamefont {Bamert}\ \emph {et~al.}(1995)\citenamefont {Bamert},
  \citenamefont {Burgess},\ and\ \citenamefont {Mohapatra}}]{Bamert:1994hb}%
  \BibitemOpen
  \bibfield  {author} {\bibinfo {author} {\bibfnamefont {P.}~\bibnamefont
  {Bamert}}, \bibinfo {author} {\bibfnamefont {C.~P.}\ \bibnamefont {Burgess}},
  \ and\ \bibinfo {author} {\bibfnamefont {R.~N.}\ \bibnamefont {Mohapatra}},\
  }\href {\doibase 10.1016/0550-3213(95)00273-U} {\bibfield  {journal}
  {\bibinfo  {journal} {Nucl. Phys.}\ }\textbf {\bibinfo {volume} {B449}},\
  \bibinfo {pages} {25} (\bibinfo {year} {1995})},\ \Eprint
  {http://arxiv.org/abs/hep-ph/9412365} {arXiv:hep-ph/9412365 [hep-ph]}
  \BibitemShut {NoStop}%
\bibitem [{\citenamefont {Hirsch}\ \emph {et~al.}(1996)\citenamefont {Hirsch},
  \citenamefont {Klapdor-Kleingrothaus}, \citenamefont {Kovalenko},\ and\
  \citenamefont {P{\"a}s}}]{Hirsch:1995in}%
  \BibitemOpen
  \bibfield  {author} {\bibinfo {author} {\bibfnamefont {M.}~\bibnamefont
  {Hirsch}}, \bibinfo {author} {\bibfnamefont {H.~V.}\ \bibnamefont
  {Klapdor-Kleingrothaus}}, \bibinfo {author} {\bibfnamefont {S.~G.}\
  \bibnamefont {Kovalenko}}, \ and\ \bibinfo {author} {\bibfnamefont
  {H.}~\bibnamefont {P{\"a}s}},\ }\href {\doibase 10.1016/0370-2693(96)00038-X}
  {\bibfield  {journal} {\bibinfo  {journal} {Phys. Lett.}\ }\textbf {\bibinfo
  {volume} {B372}},\ \bibinfo {pages} {8} (\bibinfo {year} {1996})},\ \bibinfo
  {note} {[,933(1995)]},\ \Eprint {http://arxiv.org/abs/hep-ph/9511227}
  {arXiv:hep-ph/9511227 [hep-ph]} \BibitemShut {NoStop}%
\bibitem [{\citenamefont {Blum}\ \emph {et~al.}(2018)\citenamefont {Blum},
  \citenamefont {Nir},\ and\ \citenamefont {Shavit}}]{Blum:2018ljv}%
  \BibitemOpen
  \bibfield  {author} {\bibinfo {author} {\bibfnamefont {K.}~\bibnamefont
  {Blum}}, \bibinfo {author} {\bibfnamefont {Y.}~\bibnamefont {Nir}}, \ and\
  \bibinfo {author} {\bibfnamefont {M.}~\bibnamefont {Shavit}},\ }\href
  {\doibase 10.1016/j.physletb.2018.08.022} {\bibfield  {journal} {\bibinfo
  {journal} {Phys. Lett.}\ }\textbf {\bibinfo {volume} {B785}},\ \bibinfo
  {pages} {354} (\bibinfo {year} {2018})},\ \Eprint
  {http://arxiv.org/abs/1802.08019} {arXiv:1802.08019 [hep-ph]} \BibitemShut
  {NoStop}%
\bibitem [{\citenamefont {Berezinsky}\ and\ \citenamefont
  {Valle}(1993)}]{Berezinsky:1993fm}%
  \BibitemOpen
  \bibfield  {author} {\bibinfo {author} {\bibfnamefont {V.}~\bibnamefont
  {Berezinsky}}\ and\ \bibinfo {author} {\bibfnamefont {J.~W.~F.}\ \bibnamefont
  {Valle}},\ }\href {\doibase 10.1016/0370-2693(93)90140-D} {\bibfield
  {journal} {\bibinfo  {journal} {Phys. Lett.}\ }\textbf {\bibinfo {volume}
  {B318}},\ \bibinfo {pages} {360} (\bibinfo {year} {1993})},\ \Eprint
  {http://arxiv.org/abs/hep-ph/9309214} {arXiv:hep-ph/9309214 [hep-ph]}
  \BibitemShut {NoStop}%
\bibitem [{\citenamefont {Garcia-Cely}\ and\ \citenamefont
  {Heeck}(2017)}]{Garcia-Cely:2017oco}%
  \BibitemOpen
  \bibfield  {author} {\bibinfo {author} {\bibfnamefont {C.}~\bibnamefont
  {Garcia-Cely}}\ and\ \bibinfo {author} {\bibfnamefont {J.}~\bibnamefont
  {Heeck}},\ }\href {\doibase 10.1007/JHEP05(2017)102} {\bibfield  {journal}
  {\bibinfo  {journal} {JHEP}\ }\textbf {\bibinfo {volume} {05}},\ \bibinfo
  {pages} {102} (\bibinfo {year} {2017})},\ \Eprint
  {http://arxiv.org/abs/1701.07209} {arXiv:1701.07209 [hep-ph]} \BibitemShut
  {NoStop}%
\bibitem [{\citenamefont {Brune}\ and\ \citenamefont
  {P{\"a}s}(2018)}]{Brune:2018sab}%
  \BibitemOpen
  \bibfield  {author} {\bibinfo {author} {\bibfnamefont {T.}~\bibnamefont
  {Brune}}\ and\ \bibinfo {author} {\bibfnamefont {H.}~\bibnamefont
  {P{\"a}s}},\ }\href@noop {} {\  (\bibinfo {year} {2018})},\ \Eprint
  {http://arxiv.org/abs/1808.08158} {arXiv:1808.08158 [hep-ph]} \BibitemShut
  {NoStop}%
\bibitem [{\citenamefont {P{\"a}s}\ \emph {et~al.}(1999)\citenamefont
  {P{\"a}s}, \citenamefont {Hirsch}, \citenamefont {Klapdor-Kleingrothaus},\
  and\ \citenamefont {Kovalenko}}]{Pas:1999fc}%
  \BibitemOpen
  \bibfield  {author} {\bibinfo {author} {\bibfnamefont {H.}~\bibnamefont
  {P{\"a}s}}, \bibinfo {author} {\bibfnamefont {M.}~\bibnamefont {Hirsch}},
  \bibinfo {author} {\bibfnamefont {H.~V.}\ \bibnamefont
  {Klapdor-Kleingrothaus}}, \ and\ \bibinfo {author} {\bibfnamefont {S.~G.}\
  \bibnamefont {Kovalenko}},\ }\href {\doibase 10.1016/S0370-2693(99)00330-5}
  {\bibfield  {journal} {\bibinfo  {journal} {Phys. Lett.}\ }\textbf {\bibinfo
  {volume} {B453}},\ \bibinfo {pages} {194} (\bibinfo {year}
  {1999})}\BibitemShut {NoStop}%
\bibitem [{\citenamefont {Deppisch}\ \emph {et~al.}(2012)\citenamefont
  {Deppisch}, \citenamefont {Hirsch},\ and\ \citenamefont
  {P{\"a}s}}]{Deppisch:2012nb}%
  \BibitemOpen
  \bibfield  {author} {\bibinfo {author} {\bibfnamefont {F.~F.}\ \bibnamefont
  {Deppisch}}, \bibinfo {author} {\bibfnamefont {M.}~\bibnamefont {Hirsch}}, \
  and\ \bibinfo {author} {\bibfnamefont {H.}~\bibnamefont {P{\"a}s}},\ }\href
  {\doibase 10.1088/0954-3899/39/12/124007} {\bibfield  {journal} {\bibinfo
  {journal} {J. Phys.}\ }\textbf {\bibinfo {volume} {G39}},\ \bibinfo {pages}
  {124007} (\bibinfo {year} {2012})},\ \Eprint {http://arxiv.org/abs/1208.0727}
  {arXiv:1208.0727 [hep-ph]} \BibitemShut {NoStop}%
\bibitem [{\citenamefont {Doi}\ \emph {et~al.}(1985)\citenamefont {Doi},
  \citenamefont {Kotani},\ and\ \citenamefont {Takasugi}}]{Doi:1985dx}%
  \BibitemOpen
  \bibfield  {author} {\bibinfo {author} {\bibfnamefont {M.}~\bibnamefont
  {Doi}}, \bibinfo {author} {\bibfnamefont {T.}~\bibnamefont {Kotani}}, \ and\
  \bibinfo {author} {\bibfnamefont {E.}~\bibnamefont {Takasugi}},\ }\href
  {\doibase 10.1143/PTPS.83.1} {\bibfield  {journal} {\bibinfo  {journal}
  {Prog. Theor. Phys. Suppl.}\ }\textbf {\bibinfo {volume} {83}},\ \bibinfo
  {pages} {1} (\bibinfo {year} {1985})}\BibitemShut {NoStop}%
\bibitem [{\citenamefont {Gando}\ \emph {et~al.}(2016)\citenamefont {Gando}
  \emph {et~al.}}]{KamLAND-Zen:2016pfg}%
  \BibitemOpen
  \bibfield  {author} {\bibinfo {author} {\bibfnamefont {A.}~\bibnamefont
  {Gando}} \emph {et~al.} (\bibinfo {collaboration} {KamLAND-Zen}),\ }\href
  {\doibase 10.1103/PhysRevLett.117.109903, 10.1103/PhysRevLett.117.082503}
  {\bibfield  {journal} {\bibinfo  {journal} {Phys. Rev. Lett.}\ }\textbf
  {\bibinfo {volume} {117}},\ \bibinfo {pages} {082503} (\bibinfo {year}
  {2016})},\ \bibinfo {note} {[Addendum: Phys. Rev.
  Lett.117,no.10,109903(2016)]},\ \Eprint {http://arxiv.org/abs/1605.02889}
  {arXiv:1605.02889 [hep-ex]} \BibitemShut {NoStop}%
\bibitem [{\citenamefont {Arnold}\ \emph {et~al.}(2018)\citenamefont {Arnold}
  \emph {et~al.}}]{Arnold:2018tmo}%
  \BibitemOpen
  \bibfield  {author} {\bibinfo {author} {\bibfnamefont {R.}~\bibnamefont
  {Arnold}} \emph {et~al.},\ }\href {\doibase 10.1140/epjc/s10052-018-6295-x}
  {\bibfield  {journal} {\bibinfo  {journal} {Eur. Phys. J.}\ }\textbf
  {\bibinfo {volume} {C78}},\ \bibinfo {pages} {821} (\bibinfo {year}
  {2018})},\ \Eprint {http://arxiv.org/abs/1806.05553} {arXiv:1806.05553
  [hep-ex]} \BibitemShut {NoStop}%
\bibitem [{\citenamefont {Doi}\ \emph {et~al.}(1983)\citenamefont {Doi},
  \citenamefont {Kotani}, \citenamefont {Nishiura},\ and\ \citenamefont
  {Takasugi}}]{Doi:1983wv}%
  \BibitemOpen
  \bibfield  {author} {\bibinfo {author} {\bibfnamefont {M.}~\bibnamefont
  {Doi}}, \bibinfo {author} {\bibfnamefont {T.}~\bibnamefont {Kotani}},
  \bibinfo {author} {\bibfnamefont {H.}~\bibnamefont {Nishiura}}, \ and\
  \bibinfo {author} {\bibfnamefont {E.}~\bibnamefont {Takasugi}},\ }\href
  {\doibase 10.1143/PTP.70.1353} {\bibfield  {journal} {\bibinfo  {journal}
  {Prog. Theor. Phys.}\ }\textbf {\bibinfo {volume} {70}},\ \bibinfo {pages}
  {1353} (\bibinfo {year} {1983})}\BibitemShut {NoStop}%
\bibitem [{\citenamefont {Arnold}\ \emph {et~al.}(2010)\citenamefont {Arnold}
  \emph {et~al.}}]{Arnold:2010tu}%
  \BibitemOpen
  \bibfield  {author} {\bibinfo {author} {\bibfnamefont {R.}~\bibnamefont
  {Arnold}} \emph {et~al.} (\bibinfo {collaboration} {SuperNEMO}),\ }\href
  {\doibase 10.1140/epjc/s10052-010-1481-5} {\bibfield  {journal} {\bibinfo
  {journal} {Eur. Phys. J.}\ }\textbf {\bibinfo {volume} {C70}},\ \bibinfo
  {pages} {927} (\bibinfo {year} {2010})},\ \Eprint
  {http://arxiv.org/abs/1005.1241} {arXiv:1005.1241 [hep-ex]} \BibitemShut
  {NoStop}%
\bibitem [{\citenamefont {Horoi}\ and\ \citenamefont
  {Neacsu}(2016)}]{Horoi:2015gdv}%
  \BibitemOpen
  \bibfield  {author} {\bibinfo {author} {\bibfnamefont {M.}~\bibnamefont
  {Horoi}}\ and\ \bibinfo {author} {\bibfnamefont {A.}~\bibnamefont {Neacsu}},\
  }\href {\doibase 10.1103/PhysRevD.93.113014} {\bibfield  {journal} {\bibinfo
  {journal} {Phys. Rev.}\ }\textbf {\bibinfo {volume} {D93}},\ \bibinfo {pages}
  {113014} (\bibinfo {year} {2016})},\ \Eprint
  {http://arxiv.org/abs/1511.00670} {arXiv:1511.00670 [hep-ph]} \BibitemShut
  {NoStop}%
\bibitem [{\citenamefont {Caurier}\ \emph {et~al.}(1996)\citenamefont
  {Caurier}, \citenamefont {Nowacki}, \citenamefont {Poves},\ and\
  \citenamefont {Retamosa}}]{Caurier:1996bu}%
  \BibitemOpen
  \bibfield  {author} {\bibinfo {author} {\bibfnamefont {E.}~\bibnamefont
  {Caurier}}, \bibinfo {author} {\bibfnamefont {F.}~\bibnamefont {Nowacki}},
  \bibinfo {author} {\bibfnamefont {A.}~\bibnamefont {Poves}}, \ and\ \bibinfo
  {author} {\bibfnamefont {J.}~\bibnamefont {Retamosa}},\ }\href {\doibase
  10.1103/PhysRevLett.77.1954} {\bibfield  {journal} {\bibinfo  {journal}
  {Phys. Rev. Lett.}\ }\textbf {\bibinfo {volume} {77}},\ \bibinfo {pages}
  {1954} (\bibinfo {year} {1996})},\ \Eprint
  {http://arxiv.org/abs/nucl-th/9601017} {arXiv:nucl-th/9601017 [nucl-th]}
  \BibitemShut {NoStop}%
\bibitem [{\citenamefont {Caurier}\ \emph {et~al.}(2008)\citenamefont
  {Caurier}, \citenamefont {Nowacki},\ and\ \citenamefont
  {Poves}}]{Caurier:2007qn}%
  \BibitemOpen
  \bibfield  {author} {\bibinfo {author} {\bibfnamefont {E.}~\bibnamefont
  {Caurier}}, \bibinfo {author} {\bibfnamefont {F.}~\bibnamefont {Nowacki}}, \
  and\ \bibinfo {author} {\bibfnamefont {A.}~\bibnamefont {Poves}},\ }\href
  {\doibase 10.1140/epja/i2007-10527-x} {\bibfield  {journal} {\bibinfo
  {journal} {Eur. Phys. J.}\ }\textbf {\bibinfo {volume} {A36}},\ \bibinfo
  {pages} {195} (\bibinfo {year} {2008})},\ \Eprint
  {http://arxiv.org/abs/0709.0277} {arXiv:0709.0277 [nucl-th]} \BibitemShut
  {NoStop}%
\bibitem [{\citenamefont {Macolino}(2017)}]{Macolino:2017vyd}%
  \BibitemOpen
  \bibfield  {author} {\bibinfo {author} {\bibfnamefont {C.}~\bibnamefont
  {Macolino}} (\bibinfo {collaboration} {SuperNEMO}),\ }\href {\doibase
  10.22323/1.314.0121} {\bibfield  {journal} {\bibinfo  {journal} {PoS}\
  }\textbf {\bibinfo {volume} {EPS-HEP2017}},\ \bibinfo {pages} {121} (\bibinfo
  {year} {2017})}\BibitemShut {NoStop}%
\bibitem [{\citenamefont {Gratta}(2018)}]{gratta_giorgio_2018_1286892}%
  \BibitemOpen
  \bibfield  {author} {\bibinfo {author} {\bibfnamefont {G.}~\bibnamefont
  {Gratta}},\ }\bibfield  {booktitle} {\emph {\bibinfo {booktitle} {{Talk at
  Neutrino 2018: Neutrinoless Double Beta Decay with EXO-200 and nEXO}}},\
  }\href {\doibase 10.5281/zenodo.1286892} {\  (\bibinfo {year} {2018}),\
  10.5281/zenodo.1286892}\BibitemShut {NoStop}%
\bibitem [{\citenamefont {Pati}\ and\ \citenamefont
  {Salam}(1974)}]{Pati:1974yy}%
  \BibitemOpen
  \bibfield  {author} {\bibinfo {author} {\bibfnamefont {J.~C.}\ \bibnamefont
  {Pati}}\ and\ \bibinfo {author} {\bibfnamefont {A.}~\bibnamefont {Salam}},\
  }\href {\doibase 10.1103/PhysRevD.10.275, 10.1103/PhysRevD.11.703.2}
  {\bibfield  {journal} {\bibinfo  {journal} {Phys. Rev.}\ }\textbf {\bibinfo
  {volume} {D10}},\ \bibinfo {pages} {275} (\bibinfo {year} {1974})},\ \bibinfo
  {note} {[Erratum: Phys. Rev.D11,703(1975)]}\BibitemShut {NoStop}%
\bibitem [{\citenamefont {Mohapatra}\ and\ \citenamefont
  {Pati}(1975{\natexlab{a}})}]{Mohapatra:1974hk}%
  \BibitemOpen
  \bibfield  {author} {\bibinfo {author} {\bibfnamefont {R.~N.}\ \bibnamefont
  {Mohapatra}}\ and\ \bibinfo {author} {\bibfnamefont {J.~C.}\ \bibnamefont
  {Pati}},\ }\href {\doibase 10.1103/PhysRevD.11.566} {\bibfield  {journal}
  {\bibinfo  {journal} {Phys. Rev.}\ }\textbf {\bibinfo {volume} {D11}},\
  \bibinfo {pages} {566} (\bibinfo {year} {1975}{\natexlab{a}})}\BibitemShut
  {NoStop}%
\bibitem [{\citenamefont {Mohapatra}\ and\ \citenamefont
  {Pati}(1975{\natexlab{b}})}]{Mohapatra:1974gc}%
  \BibitemOpen
  \bibfield  {author} {\bibinfo {author} {\bibfnamefont {R.~N.}\ \bibnamefont
  {Mohapatra}}\ and\ \bibinfo {author} {\bibfnamefont {J.~C.}\ \bibnamefont
  {Pati}},\ }\href {\doibase 10.1103/PhysRevD.11.2558} {\bibfield  {journal}
  {\bibinfo  {journal} {Phys. Rev.}\ }\textbf {\bibinfo {volume} {D11}},\
  \bibinfo {pages} {2558} (\bibinfo {year} {1975}{\natexlab{b}})}\BibitemShut
  {NoStop}%
\bibitem [{\citenamefont {Senjanovic}\ and\ \citenamefont
  {Mohapatra}(1975)}]{Senjanovic:1975rk}%
  \BibitemOpen
  \bibfield  {author} {\bibinfo {author} {\bibfnamefont {G.}~\bibnamefont
  {Senjanovic}}\ and\ \bibinfo {author} {\bibfnamefont {R.~N.}\ \bibnamefont
  {Mohapatra}},\ }\href {\doibase 10.1103/PhysRevD.12.1502} {\bibfield
  {journal} {\bibinfo  {journal} {Phys. Rev.}\ }\textbf {\bibinfo {volume}
  {D12}},\ \bibinfo {pages} {1502} (\bibinfo {year} {1975})}\BibitemShut
  {NoStop}%
\bibitem [{\citenamefont {Simkovic}\ \emph {et~al.}(1999)\citenamefont
  {Simkovic}, \citenamefont {Pantis}, \citenamefont {Vergados},\ and\
  \citenamefont {Faessler}}]{Simkovic:1999re}%
  \BibitemOpen
  \bibfield  {author} {\bibinfo {author} {\bibfnamefont {F.}~\bibnamefont
  {Simkovic}}, \bibinfo {author} {\bibfnamefont {G.}~\bibnamefont {Pantis}},
  \bibinfo {author} {\bibfnamefont {J.~D.}\ \bibnamefont {Vergados}}, \ and\
  \bibinfo {author} {\bibfnamefont {A.}~\bibnamefont {Faessler}},\ }\href
  {\doibase 10.1103/PhysRevC.60.055502} {\bibfield  {journal} {\bibinfo
  {journal} {Phys. Rev.}\ }\textbf {\bibinfo {volume} {C60}},\ \bibinfo {pages}
  {055502} (\bibinfo {year} {1999})},\ \Eprint
  {http://arxiv.org/abs/hep-ph/9905509} {arXiv:hep-ph/9905509 [hep-ph]}
  \BibitemShut {NoStop}%
\bibitem [{\citenamefont {Babu}\ and\ \citenamefont
  {Leung}(2001)}]{Babu:2001ex}%
  \BibitemOpen
  \bibfield  {author} {\bibinfo {author} {\bibfnamefont {K.~S.}\ \bibnamefont
  {Babu}}\ and\ \bibinfo {author} {\bibfnamefont {C.~N.}\ \bibnamefont
  {Leung}},\ }\href {\doibase 10.1016/S0550-3213(01)00504-1} {\bibfield
  {journal} {\bibinfo  {journal} {Nucl. Phys.}\ }\textbf {\bibinfo {volume}
  {B619}},\ \bibinfo {pages} {667} (\bibinfo {year} {2001})},\ \Eprint
  {http://arxiv.org/abs/hep-ph/0106054} {arXiv:hep-ph/0106054 [hep-ph]}
  \BibitemShut {NoStop}%
\bibitem [{\citenamefont {Helo}\ \emph {et~al.}(2016)\citenamefont {Helo},
  \citenamefont {Hirsch},\ and\ \citenamefont {Ota}}]{Helo:2016vsi}%
  \BibitemOpen
  \bibfield  {author} {\bibinfo {author} {\bibfnamefont {J.~C.}\ \bibnamefont
  {Helo}}, \bibinfo {author} {\bibfnamefont {M.}~\bibnamefont {Hirsch}}, \ and\
  \bibinfo {author} {\bibfnamefont {T.}~\bibnamefont {Ota}},\ }\href {\doibase
  10.1007/JHEP06(2016)006} {\bibfield  {journal} {\bibinfo  {journal} {JHEP}\
  }\textbf {\bibinfo {volume} {06}},\ \bibinfo {pages} {006} (\bibinfo {year}
  {2016})},\ \Eprint {http://arxiv.org/abs/1602.03362} {arXiv:1602.03362
  [hep-ph]} \BibitemShut {NoStop}%
\bibitem [{\citenamefont {Huang}\ and\ \citenamefont
  {Deppisch}(2015)}]{Huang:2014bva}%
  \BibitemOpen
  \bibfield  {author} {\bibinfo {author} {\bibfnamefont {W.-C.}\ \bibnamefont
  {Huang}}\ and\ \bibinfo {author} {\bibfnamefont {F.~F.}\ \bibnamefont
  {Deppisch}},\ }\href {\doibase 10.1103/PhysRevD.91.093011} {\bibfield
  {journal} {\bibinfo  {journal} {Phys. Rev.}\ }\textbf {\bibinfo {volume}
  {D91}},\ \bibinfo {pages} {093011} (\bibinfo {year} {2015})},\ \Eprint
  {http://arxiv.org/abs/1412.2027} {arXiv:1412.2027 [hep-ph]} \BibitemShut
  {NoStop}%
\bibitem [{\citenamefont {Senjanovic}(1979)}]{Senjanovic:1978ev}%
  \BibitemOpen
  \bibfield  {author} {\bibinfo {author} {\bibfnamefont {G.}~\bibnamefont
  {Senjanovic}},\ }\href {\doibase 10.1016/0550-3213(79)90604-7} {\bibfield
  {journal} {\bibinfo  {journal} {Nucl. Phys.}\ }\textbf {\bibinfo {volume}
  {B153}},\ \bibinfo {pages} {334} (\bibinfo {year} {1979})}\BibitemShut
  {NoStop}%
\bibitem [{\citenamefont {Dvali}\ \emph {et~al.}(1998)\citenamefont {Dvali},
  \citenamefont {Lazarides},\ and\ \citenamefont {Shafi}}]{Dvali:1997uq}%
  \BibitemOpen
  \bibfield  {author} {\bibinfo {author} {\bibfnamefont {G.~R.}\ \bibnamefont
  {Dvali}}, \bibinfo {author} {\bibfnamefont {G.}~\bibnamefont {Lazarides}}, \
  and\ \bibinfo {author} {\bibfnamefont {Q.}~\bibnamefont {Shafi}},\ }\href
  {\doibase 10.1016/S0370-2693(98)00145-2} {\bibfield  {journal} {\bibinfo
  {journal} {Phys. Lett.}\ }\textbf {\bibinfo {volume} {B424}},\ \bibinfo
  {pages} {259} (\bibinfo {year} {1998})},\ \Eprint
  {http://arxiv.org/abs/hep-ph/9710314} {arXiv:hep-ph/9710314 [hep-ph]}
  \BibitemShut {NoStop}%
\bibitem [{\citenamefont {Babu}\ \emph {et~al.}(2000)\citenamefont {Babu},
  \citenamefont {Pati},\ and\ \citenamefont {Wilczek}}]{Babu:1998wi}%
  \BibitemOpen
  \bibfield  {author} {\bibinfo {author} {\bibfnamefont {K.~S.}\ \bibnamefont
  {Babu}}, \bibinfo {author} {\bibfnamefont {J.~C.}\ \bibnamefont {Pati}}, \
  and\ \bibinfo {author} {\bibfnamefont {F.}~\bibnamefont {Wilczek}},\ }\href
  {\doibase 10.1016/S0550-3213(99)00589-1} {\bibfield  {journal} {\bibinfo
  {journal} {Nucl. Phys.}\ }\textbf {\bibinfo {volume} {B566}},\ \bibinfo
  {pages} {33} (\bibinfo {year} {2000})},\ \Eprint
  {http://arxiv.org/abs/hep-ph/9812538} {arXiv:hep-ph/9812538 [hep-ph]}
  \BibitemShut {NoStop}%
\bibitem [{\citenamefont {Barr}(2004)}]{Barr:2003nn}%
  \BibitemOpen
  \bibfield  {author} {\bibinfo {author} {\bibfnamefont {S.~M.}\ \bibnamefont
  {Barr}},\ }\href {\doibase 10.1103/PhysRevLett.92.101601} {\bibfield
  {journal} {\bibinfo  {journal} {Phys. Rev. Lett.}\ }\textbf {\bibinfo
  {volume} {92}},\ \bibinfo {pages} {101601} (\bibinfo {year} {2004})},\
  \Eprint {http://arxiv.org/abs/hep-ph/0309152} {arXiv:hep-ph/0309152 [hep-ph]}
  \BibitemShut {NoStop}%
\bibitem [{\citenamefont {Albright}\ and\ \citenamefont
  {Barr}(2004{\natexlab{a}})}]{Albright:2003xb}%
  \BibitemOpen
  \bibfield  {author} {\bibinfo {author} {\bibfnamefont {C.~H.}\ \bibnamefont
  {Albright}}\ and\ \bibinfo {author} {\bibfnamefont {S.~M.}\ \bibnamefont
  {Barr}},\ }\href {\doibase 10.1103/PhysRevD.69.073010} {\bibfield  {journal}
  {\bibinfo  {journal} {Phys. Rev.}\ }\textbf {\bibinfo {volume} {D69}},\
  \bibinfo {pages} {073010} (\bibinfo {year} {2004}{\natexlab{a}})},\ \Eprint
  {http://arxiv.org/abs/hep-ph/0312224} {arXiv:hep-ph/0312224 [hep-ph]}
  \BibitemShut {NoStop}%
\bibitem [{\citenamefont {Albright}\ and\ \citenamefont
  {Barr}(2004{\natexlab{b}})}]{Albright:2004ws}%
  \BibitemOpen
  \bibfield  {author} {\bibinfo {author} {\bibfnamefont {C.~H.}\ \bibnamefont
  {Albright}}\ and\ \bibinfo {author} {\bibfnamefont {S.~M.}\ \bibnamefont
  {Barr}},\ }\href {\doibase 10.1103/PhysRevD.70.033013} {\bibfield  {journal}
  {\bibinfo  {journal} {Phys. Rev.}\ }\textbf {\bibinfo {volume} {D70}},\
  \bibinfo {pages} {033013} (\bibinfo {year} {2004}{\natexlab{b}})},\ \Eprint
  {http://arxiv.org/abs/hep-ph/0404095} {arXiv:hep-ph/0404095 [hep-ph]}
  \BibitemShut {NoStop}%
\bibitem [{\citenamefont {Davidson}\ and\ \citenamefont
  {Wali}(1987)}]{Davidson:1987mh}%
  \BibitemOpen
  \bibfield  {author} {\bibinfo {author} {\bibfnamefont {A.}~\bibnamefont
  {Davidson}}\ and\ \bibinfo {author} {\bibfnamefont {K.~C.}\ \bibnamefont
  {Wali}},\ }\href {\doibase 10.1103/PhysRevLett.59.393} {\bibfield  {journal}
  {\bibinfo  {journal} {Phys. Rev. Lett.}\ }\textbf {\bibinfo {volume} {59}},\
  \bibinfo {pages} {393} (\bibinfo {year} {1987})}\BibitemShut {NoStop}%
\bibitem [{\citenamefont {Rajpoot}(1987)}]{Rajpoot:1987ji}%
  \BibitemOpen
  \bibfield  {author} {\bibinfo {author} {\bibfnamefont {S.}~\bibnamefont
  {Rajpoot}},\ }\href {\doibase 10.1103/PhysRevD.36.1479} {\bibfield  {journal}
  {\bibinfo  {journal} {Phys. Rev.}\ }\textbf {\bibinfo {volume} {D36}},\
  \bibinfo {pages} {1479} (\bibinfo {year} {1987})}\BibitemShut {NoStop}%
\bibitem [{\citenamefont {Chang}\ and\ \citenamefont
  {Mohapatra}(1987)}]{Chang:1986bp}%
  \BibitemOpen
  \bibfield  {author} {\bibinfo {author} {\bibfnamefont {D.}~\bibnamefont
  {Chang}}\ and\ \bibinfo {author} {\bibfnamefont {R.~N.}\ \bibnamefont
  {Mohapatra}},\ }\href {\doibase 10.1103/PhysRevLett.58.1600} {\bibfield
  {journal} {\bibinfo  {journal} {Phys. Rev. Lett.}\ }\textbf {\bibinfo
  {volume} {58}},\ \bibinfo {pages} {1600} (\bibinfo {year}
  {1987})}\BibitemShut {NoStop}%
\bibitem [{\citenamefont {Balakrishna}(1988)}]{Balakrishna:1987qd}%
  \BibitemOpen
  \bibfield  {author} {\bibinfo {author} {\bibfnamefont {B.~S.}\ \bibnamefont
  {Balakrishna}},\ }\href {\doibase 10.1103/PhysRevLett.60.1602} {\bibfield
  {journal} {\bibinfo  {journal} {Phys. Rev. Lett.}\ }\textbf {\bibinfo
  {volume} {60}},\ \bibinfo {pages} {1602} (\bibinfo {year}
  {1988})}\BibitemShut {NoStop}%
\bibitem [{\citenamefont {Babu}\ and\ \citenamefont
  {Mohapatra}(1989)}]{Babu:1988mw}%
  \BibitemOpen
  \bibfield  {author} {\bibinfo {author} {\bibfnamefont {K.~S.}\ \bibnamefont
  {Babu}}\ and\ \bibinfo {author} {\bibfnamefont {R.~N.}\ \bibnamefont
  {Mohapatra}},\ }\href {\doibase 10.1103/PhysRevLett.62.1079} {\bibfield
  {journal} {\bibinfo  {journal} {Phys. Rev. Lett.}\ }\textbf {\bibinfo
  {volume} {62}},\ \bibinfo {pages} {1079} (\bibinfo {year}
  {1989})}\BibitemShut {NoStop}%
\bibitem [{\citenamefont {Babu}\ and\ \citenamefont
  {Mohapatra}(1990)}]{Babu:1989rb}%
  \BibitemOpen
  \bibfield  {author} {\bibinfo {author} {\bibfnamefont {K.~S.}\ \bibnamefont
  {Babu}}\ and\ \bibinfo {author} {\bibfnamefont {R.~N.}\ \bibnamefont
  {Mohapatra}},\ }\href {\doibase 10.1103/PhysRevD.41.1286} {\bibfield
  {journal} {\bibinfo  {journal} {Phys. Rev.}\ }\textbf {\bibinfo {volume}
  {D41}},\ \bibinfo {pages} {1286} (\bibinfo {year} {1990})}\BibitemShut
  {NoStop}%
\bibitem [{\citenamefont {Brahmachari}\ \emph {et~al.}(2003)\citenamefont
  {Brahmachari}, \citenamefont {Ma},\ and\ \citenamefont
  {Sarkar}}]{Brahmachari:2003wv}%
  \BibitemOpen
  \bibfield  {author} {\bibinfo {author} {\bibfnamefont {B.}~\bibnamefont
  {Brahmachari}}, \bibinfo {author} {\bibfnamefont {E.}~\bibnamefont {Ma}}, \
  and\ \bibinfo {author} {\bibfnamefont {U.}~\bibnamefont {Sarkar}},\ }\href
  {\doibase 10.1103/PhysRevLett.91.011801} {\bibfield  {journal} {\bibinfo
  {journal} {Phys. Rev. Lett.}\ }\textbf {\bibinfo {volume} {91}},\ \bibinfo
  {pages} {011801} (\bibinfo {year} {2003})},\ \Eprint
  {http://arxiv.org/abs/hep-ph/0301041} {arXiv:hep-ph/0301041 [hep-ph]}
  \BibitemShut {NoStop}%
\bibitem [{\citenamefont {Murayama}\ and\ \citenamefont
  {Pierce}(2002)}]{Murayama:2002je}%
  \BibitemOpen
  \bibfield  {author} {\bibinfo {author} {\bibfnamefont {H.}~\bibnamefont
  {Murayama}}\ and\ \bibinfo {author} {\bibfnamefont {A.}~\bibnamefont
  {Pierce}},\ }\href {\doibase 10.1103/PhysRevLett.89.271601} {\bibfield
  {journal} {\bibinfo  {journal} {Phys. Rev. Lett.}\ }\textbf {\bibinfo
  {volume} {89}},\ \bibinfo {pages} {271601} (\bibinfo {year} {2002})},\
  \Eprint {http://arxiv.org/abs/hep-ph/0206177} {arXiv:hep-ph/0206177 [hep-ph]}
  \BibitemShut {NoStop}%
\bibitem [{\citenamefont {Thomas}\ and\ \citenamefont
  {Toharia}(2006)}]{Thomas:2005rs}%
  \BibitemOpen
  \bibfield  {author} {\bibinfo {author} {\bibfnamefont {B.}~\bibnamefont
  {Thomas}}\ and\ \bibinfo {author} {\bibfnamefont {M.}~\bibnamefont
  {Toharia}},\ }\href {\doibase 10.1103/PhysRevD.73.063512} {\bibfield
  {journal} {\bibinfo  {journal} {Phys. Rev.}\ }\textbf {\bibinfo {volume}
  {D73}},\ \bibinfo {pages} {063512} (\bibinfo {year} {2006})},\ \Eprint
  {http://arxiv.org/abs/hep-ph/0511206} {arXiv:hep-ph/0511206 [hep-ph]}
  \BibitemShut {NoStop}%
\bibitem [{\citenamefont {Thomas}\ and\ \citenamefont
  {Toharia}(2007)}]{Thomas:2006gr}%
  \BibitemOpen
  \bibfield  {author} {\bibinfo {author} {\bibfnamefont {B.}~\bibnamefont
  {Thomas}}\ and\ \bibinfo {author} {\bibfnamefont {M.}~\bibnamefont
  {Toharia}},\ }\href {\doibase 10.1103/PhysRevD.75.013013} {\bibfield
  {journal} {\bibinfo  {journal} {Phys. Rev.}\ }\textbf {\bibinfo {volume}
  {D75}},\ \bibinfo {pages} {013013} (\bibinfo {year} {2007})},\ \Eprint
  {http://arxiv.org/abs/hep-ph/0607285} {arXiv:hep-ph/0607285 [hep-ph]}
  \BibitemShut {NoStop}%
\bibitem [{\citenamefont {Abel}\ and\ \citenamefont
  {Page}(2006)}]{Abel:2006hr}%
  \BibitemOpen
  \bibfield  {author} {\bibinfo {author} {\bibfnamefont {S.}~\bibnamefont
  {Abel}}\ and\ \bibinfo {author} {\bibfnamefont {V.}~\bibnamefont {Page}},\
  }\href {\doibase 10.1088/1126-6708/2006/05/024} {\bibfield  {journal}
  {\bibinfo  {journal} {JHEP}\ }\textbf {\bibinfo {volume} {05}},\ \bibinfo
  {pages} {024} (\bibinfo {year} {2006})},\ \Eprint
  {http://arxiv.org/abs/hep-ph/0601149} {arXiv:hep-ph/0601149 [hep-ph]}
  \BibitemShut {NoStop}%
\bibitem [{\citenamefont {Cerdeno}\ \emph {et~al.}(2006)\citenamefont
  {Cerdeno}, \citenamefont {Dedes},\ and\ \citenamefont
  {Underwood}}]{Cerdeno:2006ha}%
  \BibitemOpen
  \bibfield  {author} {\bibinfo {author} {\bibfnamefont {D.~G.}\ \bibnamefont
  {Cerdeno}}, \bibinfo {author} {\bibfnamefont {A.}~\bibnamefont {Dedes}}, \
  and\ \bibinfo {author} {\bibfnamefont {T.~E.~J.}\ \bibnamefont {Underwood}},\
  }\href {\doibase 10.1088/1126-6708/2006/09/067} {\bibfield  {journal}
  {\bibinfo  {journal} {JHEP}\ }\textbf {\bibinfo {volume} {09}},\ \bibinfo
  {pages} {067} (\bibinfo {year} {2006})},\ \Eprint
  {http://arxiv.org/abs/hep-ph/0607157} {arXiv:hep-ph/0607157 [hep-ph]}
  \BibitemShut {NoStop}%
\bibitem [{\citenamefont {Gu}\ and\ \citenamefont {He}(2006)}]{Gu:2006dc}%
  \BibitemOpen
  \bibfield  {author} {\bibinfo {author} {\bibfnamefont {P.-H.}\ \bibnamefont
  {Gu}}\ and\ \bibinfo {author} {\bibfnamefont {H.-J.}\ \bibnamefont {He}},\
  }\href {\doibase 10.1088/1475-7516/2006/12/010} {\bibfield  {journal}
  {\bibinfo  {journal} {JCAP}\ }\textbf {\bibinfo {volume} {0612}},\ \bibinfo
  {pages} {010} (\bibinfo {year} {2006})},\ \Eprint
  {http://arxiv.org/abs/hep-ph/0610275} {arXiv:hep-ph/0610275 [hep-ph]}
  \BibitemShut {NoStop}%
\bibitem [{\citenamefont {Gu}\ \emph {et~al.}(2008)\citenamefont {Gu},
  \citenamefont {He},\ and\ \citenamefont {Sarkar}}]{Gu:2007mc}%
  \BibitemOpen
  \bibfield  {author} {\bibinfo {author} {\bibfnamefont {P.-H.}\ \bibnamefont
  {Gu}}, \bibinfo {author} {\bibfnamefont {H.-J.}\ \bibnamefont {He}}, \ and\
  \bibinfo {author} {\bibfnamefont {U.}~\bibnamefont {Sarkar}},\ }\href
  {\doibase 10.1016/j.physletb.2007.11.061} {\bibfield  {journal} {\bibinfo
  {journal} {Phys. Lett.}\ }\textbf {\bibinfo {volume} {B659}},\ \bibinfo
  {pages} {634} (\bibinfo {year} {2008})},\ \Eprint
  {http://arxiv.org/abs/0709.1019} {arXiv:0709.1019 [hep-ph]} \BibitemShut
  {NoStop}%
\bibitem [{\citenamefont {Gu}\ \emph {et~al.}(2007)\citenamefont {Gu},
  \citenamefont {He},\ and\ \citenamefont {Sarkar}}]{Gu:2007mi}%
  \BibitemOpen
  \bibfield  {author} {\bibinfo {author} {\bibfnamefont {P.-H.}\ \bibnamefont
  {Gu}}, \bibinfo {author} {\bibfnamefont {H.-J.}\ \bibnamefont {He}}, \ and\
  \bibinfo {author} {\bibfnamefont {U.}~\bibnamefont {Sarkar}},\ }\href
  {\doibase 10.1088/1475-7516/2007/11/016} {\bibfield  {journal} {\bibinfo
  {journal} {JCAP}\ }\textbf {\bibinfo {volume} {0711}},\ \bibinfo {pages}
  {016} (\bibinfo {year} {2007})},\ \Eprint {http://arxiv.org/abs/0705.3736}
  {arXiv:0705.3736 [hep-ph]} \BibitemShut {NoStop}%
\end{thebibliography}%

\end{document}